\documentclass[a4paper,fleqn,usenatbib,useAMS]{mnras}
\usepackage{graphicx}   
\usepackage{color}
\usepackage{lscape}
\usepackage{txfonts}
\usepackage{amsfonts}

\title[X-ray emission in low-level accreting NSs]
{A systematic study of the advection-dominated accretion flow for the origin of 
the X-ray emission in weakly magnetized low-level accreting neutron stars}
\author[Erlin Qiao and B.F. Liu]{Erlin Qiao $^{1,2}$\thanks{E-mail:
qiaoel@nao.cas.cn} and B.F. Liu $^{1,2}$\\
$^{1}$Key Laboratory of Space Astronomy and Technology, National Astronomical Observatories, Chinese Academy of
Sciences, Beijing 100012, China \\
$^{2}$School of Astronomy and
Space Sciences, University of Chinese Academy of Sciences, 19A Yuquan Road, Beijing 100049, China\\}

\date{Accepted XXX. Received YYY; in original form ZZZ}
\pubyear{2019}

\begin{document}
\label{firstpage}
\pagerange{\pageref{firstpage}--\pageref{lastpage}}
\maketitle
\begin{abstract}
Observationally, the X-ray spectrum ($0.5-10$ keV) of low-level accreting neutron stars (NSs) 
($L_{\rm 0.5-10\rm keV}\lesssim 10^{36}\ \rm erg \ s^{-1}$) can generally be well fitted by the 
model with two components, i.e, a thermal soft X-ray component plus a power-law component. 
Meanwhile, the fractional contribution of the power-law luminosity $\eta$ 
($\eta\equiv L^{\rm power\ law}_{\rm 0.5-10\rm keV}/L_{\rm 0.5-10\rm keV}$) varies with the 
X-ray luminosity $L_{\rm 0.5-10\rm keV}$. In this paper, we systematically investigate  
the origin of such X-ray emission within the framework of the advection-dominated 
accretion flow (ADAF) around a weakly magnetized NS, in which the thermal soft X-ray component 
arises from the surface of the NS and the power-law component arises from the ADAF itself. 
We test the effects of the viscosity parameter $\alpha$ in the ADAF and thermalized parameter $f_{\rm th}$ 
(describing the fraction of the ADAF energy released at the surface of the NS as thermal emission) 
on the relation of $\eta$ versus $L_{\rm 0.5-10\rm keV}$.  
It is found that $\eta$ is nearly a constant ($\sim$ zero) with $L_{\rm 0.5-10\rm keV}$
for different $\alpha$ with $f_{\rm th}=1$, which is inconsistent with observations.
Meanwhile, it is found that a change of $f_{\rm th}$ can significantly change the relation of 
$\eta$ versus $L_{\rm 0.5-10\rm keV}$. 
By comparing with a sample of non-pulsating NS-low mass X-ray binaries probably dominated by low-level 
accretion onto NSs, it is found that a small value of $f_{\rm th} \lesssim 0.1$ is needed to match the 
observed range of $\eta \gtrsim 10\%$ in the diagram of $\eta$ versus $L_{\rm 0.5-10\rm keV}$.
Finally, we argue that the small value of $f_{\rm th} \lesssim 0.1$ implies that the 
radiative efficiency of NSs with an ADAF accretion may not be as high as the predicted result 
previously of $\epsilon \sim {\dot M GM\over R_{*}}/{\dot M c^2}\sim 0.2$ despite the existence 
of the hard surface.
\end{abstract}



\section{Introduction}
Most of the neutron star low-mass X-ray binaries (NS-LMXBs) are soft X-ray transients,
which spend most of their time in the quiescent state, and are generally discovered when 
they first go into outburst \citep[][for review]{Campana1998}.
NS-LMXBs can be divided into two distinct classes, i.e., the atoll-type NSs and the Z-type NSs, 
with the name derived from the shape they trace in the color-color diagram during the outburst and decay 
\citep[][]{Hasinger1989}. 
Atoll-type NSs are seen in the luminosity range from $\lesssim 10^{-3} L_{\rm Edd}$ 
up to $\sim L_{\rm Edd}$ (with $L_{\rm Edd}=1.26\times 10^{38} M/M_{\odot}\ \rm {erg\ s^{-1}}$), 
showing two distinct states, i.e., the soft state with a relatively higher mass accretion rate and the hard 
state with a relatively lower mass accretion rate as in black hole (BH) X-ray binaries according to 
the spectral and timing features. Z-type NSs accrete at or near the Eddington accretion rate with the 
luminosity greater than $\sim 0.5 L_{\rm Edd}$ \citep[e.g.][]{Done2003}.
The non-pulsating NS-LMXBs generally have a low magnetic field ($B\lesssim 10^{8-9} \rm G$), so it is 
expected that the magnetic field does not affect the dynamics of the accretion flow. 
When the 
atoll-type NSs are in the soft state, the accretion is believed to be dominated by the 
optically thick, geometrically thin, cool disc \citep[][]{Shakura1973}. 
Theoretically, in the case of Newtonian mechanics, one half of the gravitational energy 
will be released in the disc as the multi-color blackbody spectrum with a typical 
temperature $\sim 10^7$ K, and the other half of the gravitational energy will be released 
in a thin boundary layer between the accretion disc and the surface of the NS 
\citep[][]{Popham1992,Inogamov1999,Popham2001,Frank2002}.
In the case of General Relativity, the energy released in the boundary layer is even larger, 
roughly twice than that of the disc \citep[][]{Syunyaev1986,Sibgatullin2000}. 
The boundary layer is optically thick, and the temperature of the boundary layer is 
several times higher than that of the inner disc, producing a blackbody spectrum with a 
higher temperature \citep[][for review]{Gilfanov2014}.

When the atoll-type NSs are in the hard state, theoretically, the accretion flow is suggested to 
be dominated by the so-called advection-dominated accretion flow (ADAF)\citep[][]{Narayan1995b,Esin1997}. 
The properties of the ADAF around a NS are different from that of a BH due to the existence of a hard surface 
of NSs, which is critically examined by considering the effects of the boundary layer between the ADAF and 
the central NS \citep[][]{Medvedev2001,Medvedev2004}.
\citet[][]{Qiao2018b} investigated the properties of the ADAF around a weakly magnetized NS
in the framework of the self-similar solution of the ADAF. 
In \citet[][]{Qiao2018b}, the authors assumed that a fraction, $f_{\rm th}$,
of the energy in the ADAF (including the internal energy and the radial kinetic energy) 
transferred onto the surface of the NS is thermalized at the surface of the NS as the soft 
photons to be scattered in the ADAF itself. Then the authors self-consistently calculate the structure
of the ADAF and the corresponding emergent spectrum by considering the radiative coupling
of the soft photons from the surface of the NS and the ADAF itself. Theoretically, it can be seen that 
if the authors take the viscosity parameter $\alpha=0.3$ and $f_{\rm th}=1$, 
$\sim 60\%$ of the  gravitational energy will be released at the surface of the NS in the form 
of the blackbody emission \citep[][]{Qiao2018b}.

Observationally, the X-ray spectrum of NS-LMXBs in the hard state has been well studied for 
the X-ray luminosity above $\sim 10^{36}\ \rm erg \ s^{-1}$. While for low-level accreting
NSs, i.e., the X-ray luminosity below $\sim 10^{36}\ \rm erg \ s^{-1}$, especially the X-ray luminosity 
below $\sim 10^{34}\ \rm erg \ s^{-1}$ (generally defined as the quiescent state), the X-ray spectrum is 
relatively less well understood \citep[e.g.][]{DAngelo2015,Chakrabarty2014,Wijnands2015}.
The X-ray spectrum in the range of $0.5-10$ keV for low-level accreting NSs 
can generally be well fitted with the model with two components, i.e., a thermal soft X-ray component 
plus a power-law component \citep[e.g.][]{Jonker2004,Campana2008a,Campana2008b,Fridriksson2010,Fridriksson2011,
ArmasPadilla2013b,Degenaar2013,Campana2014,Parikh2017,Vats2018}. 
For certain low-level accreting NSs, the temperature of the soft X-ray component decreases with 
decreasing the X-ray luminosity during the decay after the outburst
\citep[][]{ArmasPadilla2013c,Degenaar2013,Bahramian2014}.
Some studies show that the fractional contribution of the power-law luminosity 
$\eta$ ($\eta=L^{\rm power\ law}_{\rm 0.5-10\rm keV}/L_{\rm 0.5-10\rm keV}$) is in a relatively 
narrow range of $\eta \sim 40-60\%$ \citep[][]{Cackett2010,Fridriksson2011,ArmasPadilla2013b,
Campana2014,Homan2014,Bahramian2014}. While some studies show that $\eta$ can be in a wide range,
e.g., $\eta$ is in the range of $10-40\%$ for Aql X-1, Cen X-4 
\citep[][]{Asai1996,Campana2000,Rutledge2000}, 4U 2129+47 \citep[][]{Nowak2002}, and 
even an extreme value of $\eta \sim 100\%$ for SAX J1808.4-3658 \citep[][]{Campana2002}. 
\citet[][]{Jonker2004} compiled a sample composed of 12 NS soft X-ray transients with the distance 
well-determined, the authors showed that there seems to be a positive correlation between $\eta$ 
and $L_{\rm 0.5-10\rm keV}$ for $L_{\rm 0.5-10\rm keV} \gtrsim 2\times 10^{33} \rm \ erg \ s^{-1}$, 
below which there seems to be an anti-correlation between $\eta$ and $L_{\rm 0.5-10\rm keV}$ for
some sources.

So far, although a great of efforts have been made for investigating the X-ray spectrum in the range 
of 0.5-10 keV for low-level accreting NSs, the physical origin of the X-ray emission including both 
the thermal soft X-ray component and the power-law component, especially the relative strength between  
the thermal soft X-ray component and the power-law component is still under debate. 
Several models have been proposed for the origin of the X-ray emission 
in the hard state and the quiescent state of NS-LMXBs. The thermal soft X-ray component could be 
from the crust cooling of the NS, which was heated during the last outburst 
\citep[e.g.][]{Brown1998,Campana2000,Rutledge1999,Rutledge2001a,Rutledge2001b}, or 
could be related to the low-level accretion onto the surface of the NS
\citep[e.g.][]{Zampieri1995,Campana1997,Cackett2011,Bernardini2013,Qiao2018b}.
The power-law component could be related to the magnetic field of the NS, e.g., for SAX J1808.4-3658 
\citep[][]{Campana2002}, or also could be related to the low-level accretion onto the 
surface of the NS, e.g., for Cen X-4 \citep[][]{DAngelo2015,Chakrabarty2014}.  

In this paper, we systematically investigate the origin of 
the X-ray emission in low-level accreting NSs within the framework of the self-similar solution of 
the ADAF as in \citet[][]{Qiao2018b}, in which the thermal soft X-ray component arises  
from the surface of the NS, and the power-law component arises from the ADAF itself.
Additionally, in this paper, we update the calculation of \citet[][]{Qiao2018b} with the effects 
of the NS spin considered. We consider that the internal energy and the radial kinetic energy of the 
ADAF as in \citet[][]{Qiao2018b}, as well as the rotational energy of the ADAF are transferred onto the 
surface of the NS. The rotational energy of the ADAF transferred onto the 
surface of the NS is dependent on the NS spin \citep[][]{Gilfanov2014}. We assume that a fraction, 
$f_{\rm th}$, of the total energy (including the internal energy, the radial kinetic energy
and the rotational energy of the ADAF) transferred onto the surface of the NS is thermalized at the surface of the NS 
as the soft photons to be scattered in the ADAF. Then we self-consistently calculate the structure 
and the corresponding emergent spectrum of the ADAF by considering the radiative coupling   
between the soft photons from the surface of the NS and the ADAF itself. 
Based on the theoretical emergent spectra, we investigate the relation between $\eta$ 
and $L_{\rm 0.5-10\rm keV}$. Specifically, we test the effects of the 
viscosity parameter $\alpha$ in the ADAF and the thermalized parameter $f_{\rm th}$ on the 
relation between $\eta$ and $L_{\rm 0.5-10\rm keV}$. It is found that the effect of $\alpha$ on the  
relation between $\eta$ and $L_{\rm 0.5-10\rm keV}$ is very little, and can nearly be neglected. 
As an example, $\eta$ is nearly a constant ($\sim$ zero) with $L_{\rm 0.5-10\rm keV}$ for different $\alpha$ with
$f_{\rm th}=1$. Meanwhile, we find that a change of $f_{\rm th}$ can significantly change the relation between 
$\eta$ and  $L_{\rm 0.5-10\rm keV}$. 
By comparing with a sample of non-pulsating NS-LMXBs probably dominated by low-level accretion onto NSs, we find that 
a small value of $f_{\rm th} \lesssim 0.1$ can match the observed range of $\eta \gtrsim 10\%$
in the diagram of $\eta$ versus $L_{\rm 0.5-10\rm keV}$.
The derived small value of $f_{\rm th}\lesssim 0.1$ suggests that the radiative efficiency of 
a weakly magnetized NS with an ADAF accretion
may not be as high as $\epsilon \sim {\dot M GM\over R_{*}}/{\dot M c^2}\sim 0.2$ 
as predicted previously. The model is briefly introduced in Section 2. The numerical results are shown in Section 3. 
The discussions are in Section 4 and the conclusions are in Section 5.

\section{The model}\label{model}
We calculate the structure of the ADAF around a weakly magnetized NS within the framework 
of the self-similar solution of the ADAF as in \citet[][]{Qiao2018b}. In this paper, we 
consider that the internal energy and the radial kinetic energy of the ADAF as in \citet[][]{Qiao2018b}, 
as well as the rotational energy of the ADAF are transferred onto the surface of the NS. 
For clarity, we list the expression of the internal energy and the radial kinetic energy of 
the ADAF transferred onto the surface of the NS per second, $L_{*}^{'}$ (see also equation (9) in 
\citet[][]{Qiao2018b}), which is as follows,  
\begin{equation}\label{soft_1}
\begin{array}{l}
L_{*}^{'} = 4\pi R_{*}H(R_{*})|{\rm v}(R_{*})|\left[U(R_{*})+{1\over2}\rho(R_{*}){\rm v}^2(R_{*})\right], 
\end{array}
\end{equation}
where $R_{*}$ is the radius of the NS, and $U(R_{*})$ is the internal energy of the gas
at $R_{*}$, $H(R_{*})$ is the scaleheight of the gas at $R_{*}$, $\rho(R_{*})$ and  ${\rm v}(R_{*})$ 
are the density of the gas and the radial velocity of the gas at $R_{*}$ respectively.
The rotational energy of the ADAF transferred onto the 
surface of the NS per second, $L_{\rm bl}$, ($L_{\rm bl}$ occurs actually in a thin boundary layer 
between the ADAF and the surface of the NS) can be expressed as,  
\begin{eqnarray}\label{soft_2}
L_{\rm bl}=2\pi^2 \dot M R_{*}^2 (\nu_{*}-\nu_{\rm NS})^2,
\end{eqnarray}
where $\nu_{\rm *}$ is the rotational frequency of the accretion flow at $R_{*}$
(with $\nu_{\rm *}=\Omega_{\rm *}/{2\pi}$, and $\Omega_{\rm *}$ being the 
angular velocity of the ADAF at $R_{*}$), and $\nu_{\rm NS}$ is the rotational frequency of the NS.
Then the total energy of the ADAF transferred onto the surface of the NS per second, $L_{*}$,  
can be expressed as the sum of $L_{*}^{'}$ and $L_{\rm bl}$,  i.e., 
\begin{eqnarray}\label{equ:soft_L}
L_{*} = L_{*}^{'}+ L_{\rm bl}.
\end{eqnarray} 
As in \citet[][]{Qiao2018b}, we assume that a fraction, $f_{\rm th}$, of the energy, 
$L_{*}$, is thermalized at the surface of the NS as the soft photons to be scattered in the 
ADAF to self-consistently calculate the structure of the ADAF.
In this case, if the radiation from the surface of the NS is isotropic, the effective temperature 
of the radiation $T_{*}$ can then be given as,
\begin{equation}\label{equ:T}
\begin{array}{l}
T_{*}=\bigl({L_{*}f_{\rm th}\over {4\pi R_{*}^{2}\sigma}}\bigr)^{1/4}, 
\end{array}
\end{equation}
where $\sigma$ is the Stefan-Boltzmann constant. 
Further, replacing equation (9) in \citet[][]{Qiao2018b} by equation \ref{equ:soft_L} in this paper, 
we can calculate the ion temperature $T_{\rm i}$, electron temperature $T_{\rm e}$ and the
advected fraction of the viscously dissipated energy $f$  as in \citet[][]{Qiao2018b} 
by specifying the model parameters, i.e., the NS mass $m$ ($m=M/M_{\odot}$), 
the NS radius $R_{*}$, the mass accretion rate $\dot m$\footnote{In this paper, a constant 
$\dot m$ as a function of radius is assumed, which has been supported by the recent numerical 
simulations of the ADAF around a weakly magnetized NS \citep[][]{Bu2019}.} (with $\dot m=\dot M/\dot M_{\rm Edd}$,  
$\dot M_{\rm Edd}=L_{\rm Edd}/{0.1c^2}$=$1.39 \times 10^{18} M/M_{\rm \odot} \rm \ g \ s^{-1}$),
the viscosity parameter $\alpha$, the magnetic parameter $\beta$
(with magnetic pressure $p_{\rm m}={B^2/{8\pi}}=(1-\beta)p_{\rm tot}$,
$p_{\rm tot}=p_{\rm gas}+p_{\rm m}$), $f_{\rm th}$ describing the fraction of the total energy 
the ADAF transferred onto the surface of the NS to be thermalized as the blackbody emission, 
and the rotational frequency of the NS $\nu_{\rm NS}$.
The other quantities of ADAF, such as the Compton scattering optical depth in the vertical direction
$\tau_{\rm es}$, the Compton $y$-parameter (defined as 
$y={{4kT_{\rm e}}\over {m_{\rm e}c^2}} \tau_{\rm es}$, with $\tau_{\rm es}<1$), 
and the angular velocity $\Omega$ of the ADAF can be calculated with the formulae derived by 
the self-similar solution of the ADAF accordingly [e.g., equation (4) in \citet[][]{Qiao2018b}].
In this paper, we set the NS mass $m=1.4$, and the NS radius $R_{*}=12.5$ km as in \citet[][]{Qiao2018b}.
In the ADAF solution, the magnetic field is often relatively weak,  
as suggested by the magnetohydrodynamic simulations \citep[][for review]{Yuan2014}.
Throughout this paper, we set $\beta=0.95$. 
We test the effect of the rotational frequency of the NS $\nu_{\rm NS}$ 
on the structure of the ADAF in Section \ref{spin}, it is found that the effect of  
$\nu_{\rm NS}$ on the structure of the ADAF can nearly be neglected. 
Throughout this paper, we set $\nu_{\rm NS}=0$. So, we have two parameters left, i.e., 
$\alpha$ and $f_{\rm th}$. Finally, we calculate the emergent spectrum of the ADAF around
a NS with the method of multi-scattering of soft photons in the hot gas 
\citep[e.g.][]{Manmoto1997,Qiao2010,Qiao2013,Qiao2018b}.

\section{The results}
\subsection{The effect of $\alpha$}\label{Sec:alpha}
\begin{figure*}
\includegraphics[width=85mm,height=60mm,angle=0.0]{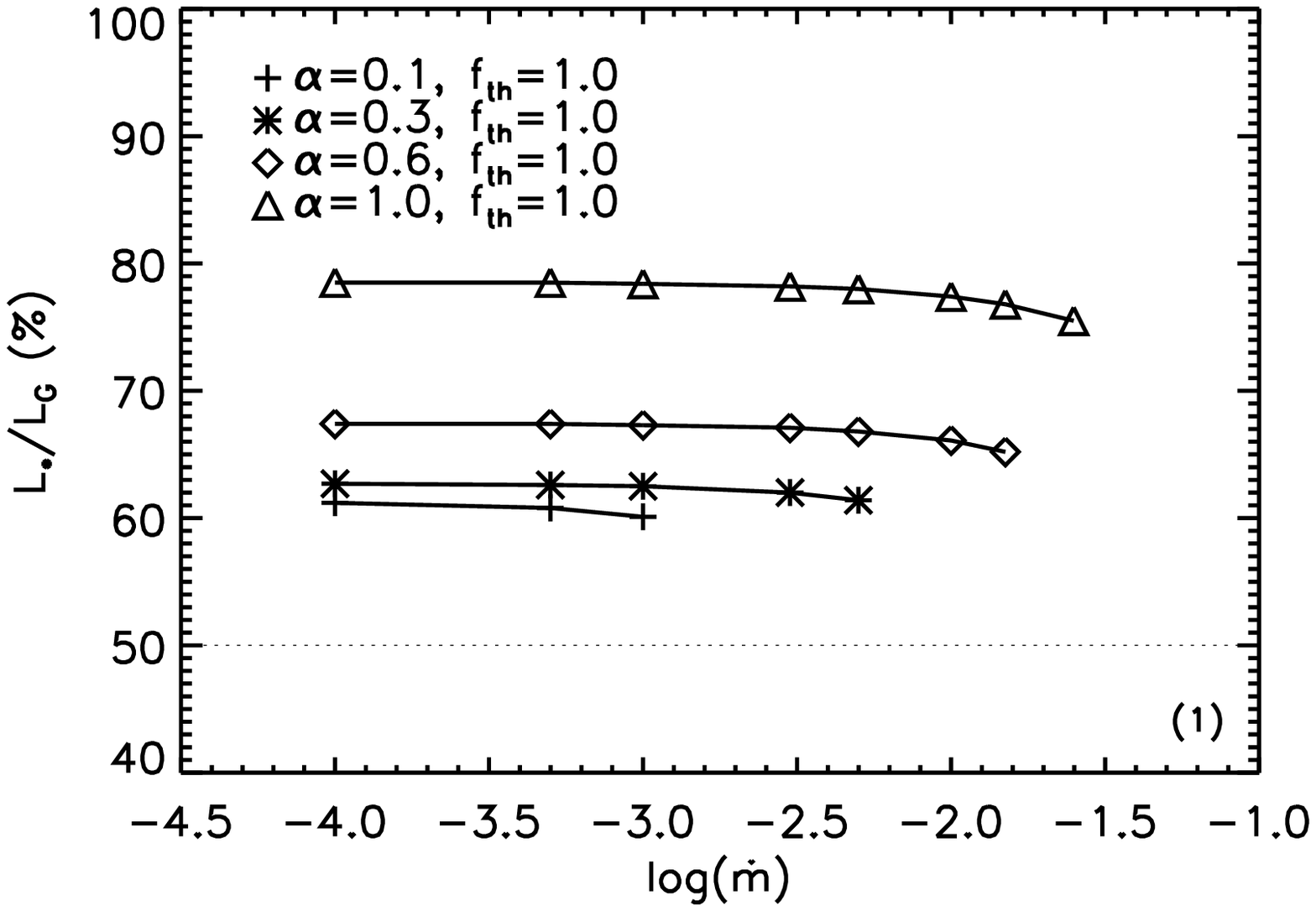}
\includegraphics[width=85mm,height=60mm,angle=0.0]{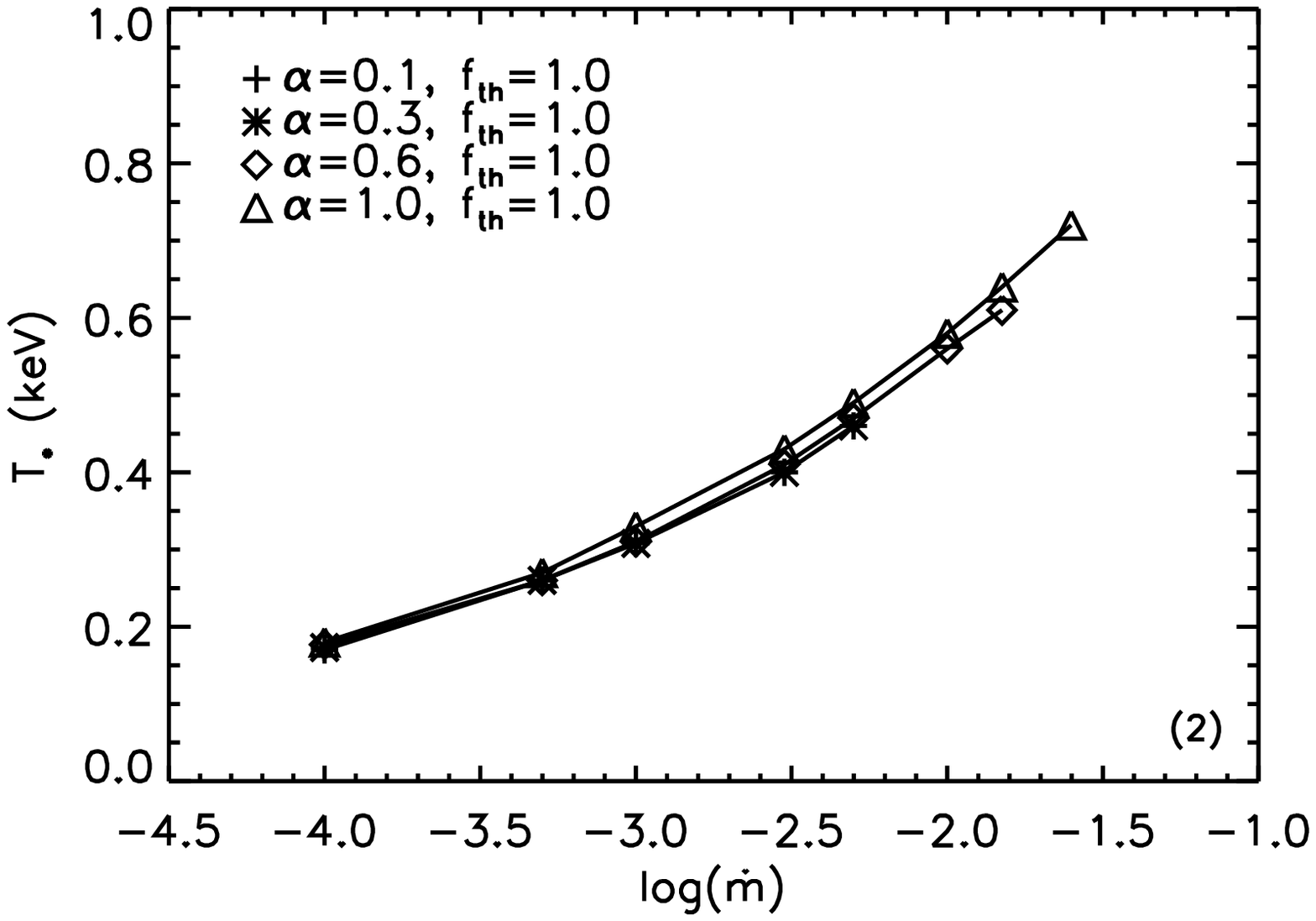}
\includegraphics[width=85mm,height=60mm,angle=0.0]{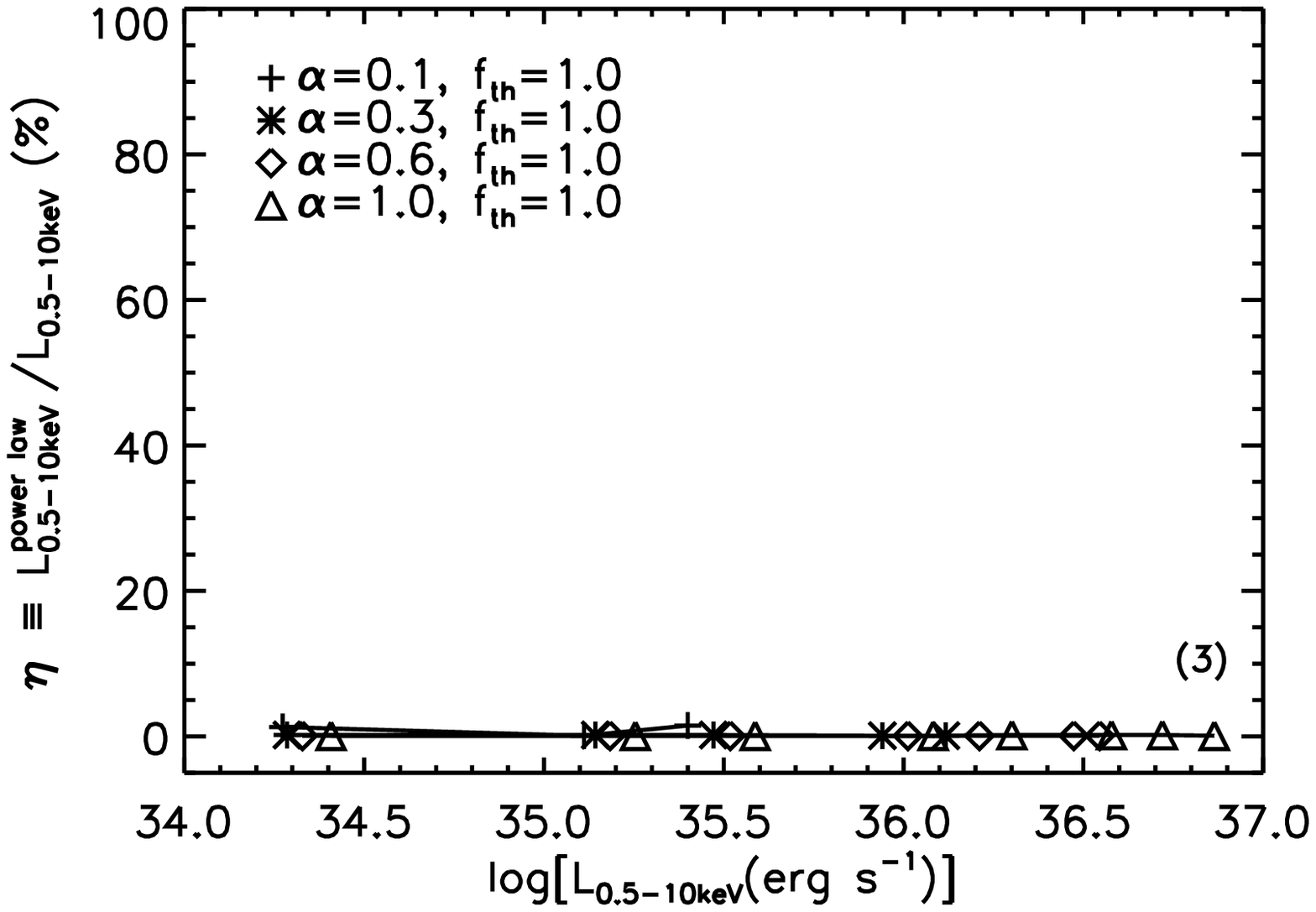}
\includegraphics[width=85mm,height=60mm,angle=0.0]{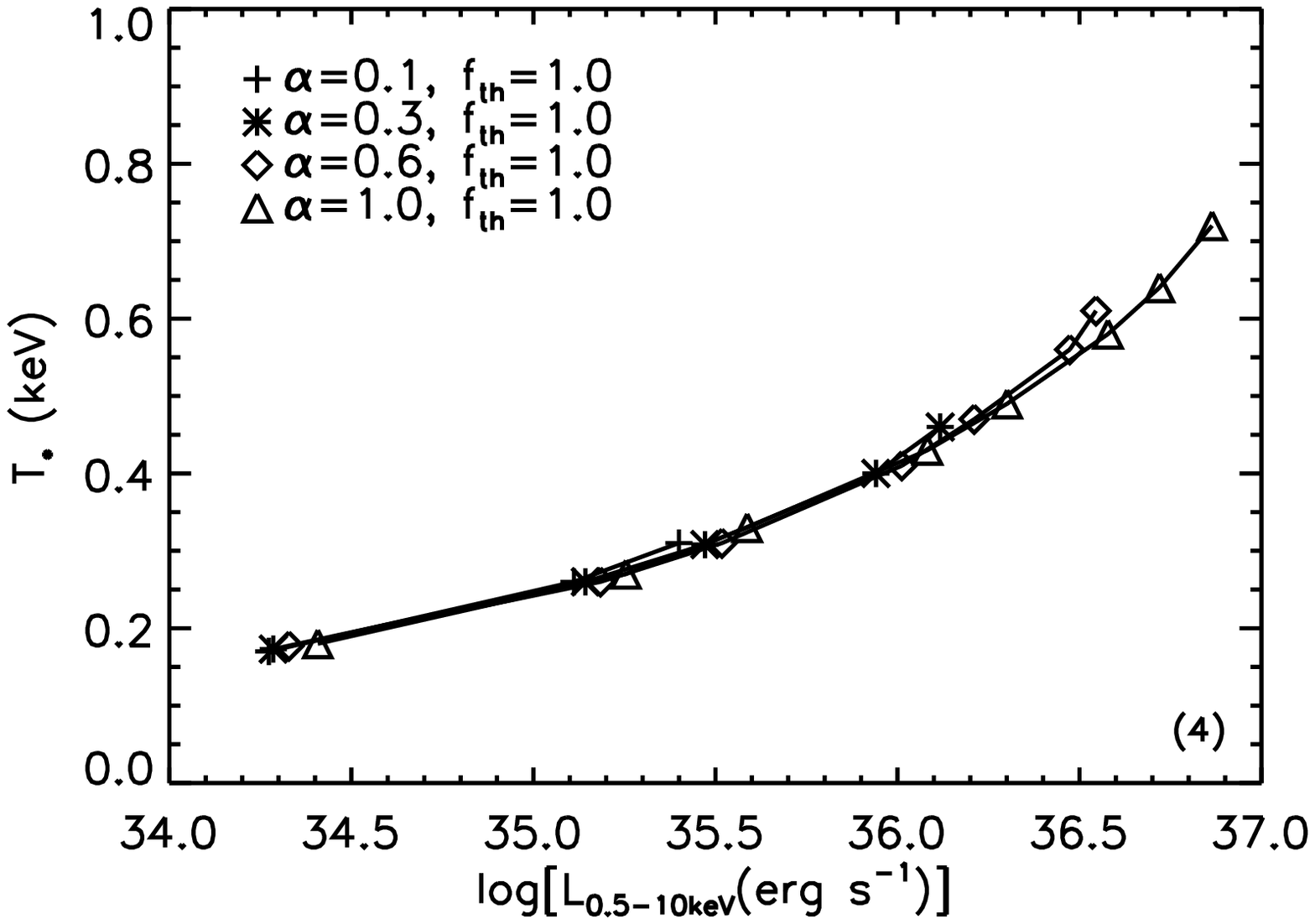}
\caption{\label{f:alpha}Panel (1): Ratio of the energy transferred onto
the surface of the NS per second $L_{*}$ to the accretion luminosity $L_{\rm G}$, 
$L_{*}/L_{\rm G}$, as a function of $\dot m$.
Panel (2): Effective temperature at the surface of the NS $T_{*}$ as a function of $\dot m$.  
Panel (3): Fractional contribution of the power-law luminosity 
$\eta\equiv L^{\rm power\ law}_{\rm 0.5-10\rm keV}/L_{\rm 0.5-10\rm keV}$ as a function of the X-ray luminosity  
$L_{\rm 0.5-10\rm keV}$.
Panel (4): Effective temperature at the surface of the NS $T_{*}$ as a function of the 
X-ray luminosity  $L_{\rm 0.5-10\rm keV}$.
}
\end{figure*}

In the panel (1) of Fig. \ref{f:alpha}, we plot the ratio of the energy transferred onto
the surface of the NS per second $L_{*}$ to the accretion luminosity $L_{\rm G}$ 
(defined as $L_{\rm G}={GM\dot M}/R_{*}$), 
$L_{*}/L_{\rm G}$, as a function of $\dot m$ for different $\alpha$ with $f_{\rm th}=1.0$.
It can be seen that $L_{*}/L_{\rm G}$ increases very slightly with decreasing $\dot m$ 
for $\alpha=0.1, 0.3, 0.6$ and $1.0$ respectively.
One can refer to Table \ref{T:alpha_effect} for the detailed numerical results \footnote{Since there is a 
critical mass accretion rate for the ADAF solution around a NS, i.e., 
$\dot M_{\rm crit} \sim 0.1\alpha^2 \dot M_{\rm Edd}$, above which the ADAF solution cannot exist. 
As we can see, $\dot M_{\rm crit}$ is very sensitive to the viscosity parameter $\alpha$, so
we roughly take $\dot m=1.0\times 10^{-3}$ for $\alpha=0.1$,  
$\dot m=5.0\times 10^{-3}$ for $\alpha=0.3$, 
$\dot m=1.5\times 10^{-2}$ for $\alpha=0.6$, and  
$\dot m=2.5\times 10^{-3}$ for $\alpha=1.0$ as the upper limits of the mass accretion rate respectively
in our calculations.}.
Meanwhile, it is clear that, for a fixed $\dot m$, the derived $L_{*}/L_{\rm G}$ with a bigger 
value of $\alpha$ is systematically higher than that of with a smaller value of $\alpha$,
which can be understood as follows. Generally, as we know the luminosity of the ADAF around a BH 
can be expressed as $L_{\rm ADAF} \propto \dot m^{2} \alpha^{-2}$ \citep[e.g.][]{Mahadevan1997}, 
so for a fixed $\dot m$, an increase the value of $\alpha$ will result in a decrease of the 
radiation of the ADAF, which {in turn} means that the energy transferred into the event horizon 
of the BH will increase with increasing $\alpha$. As far as NSs, feedback exists between the 
energy transferred onto the surface of the NS and the ADAF itself.
As a whole, due to the hot nature of the ADAF around both BHs and NSs, the radiation of
the ADAF itself will decrease with increasing $\alpha$, which means that the energy transferred onto 
the surface of the NS will increase with increasing $\alpha$ for a fixed $\dot m$, as can be seen in
the panel (1) of Fig. \ref{f:alpha}. 
In the panel (2) of Fig. \ref{f:alpha}, we plot the temperature at the surface of the NS $T_{*}$ as
a function of $\dot m$ for different $\alpha$ with $f_{\rm th}=1.0$.
It is clear that $T_{*}$ decreases with decreasing $\dot m$ for $\alpha=0.1, 0.3, 0.6$ and $1.0$ 
respectively. Meanwhile, it can be seen that the effect of $\alpha$ on the relation between 
$T_{*}$ and $\dot m$ is very little, which can be understood as, although roughly there is an increase
of $\sim 20\%$ for the value of $L_{*}/L_{\rm G}$ for taking $\alpha=1.0$ compared with taking 
$\alpha=0.1$ at a fixed $\dot m$,  according to equation \ref {equ:T}, 
the increase of $T_{*}$  is only $(1+20\%)^{1/4}-1 \sim 4.7\%$.
One can refer to Table \ref{T:alpha_effect} for the detailed numerical results.

\begin{figure*}
\includegraphics[width=85mm,height=60mm,angle=0.0]{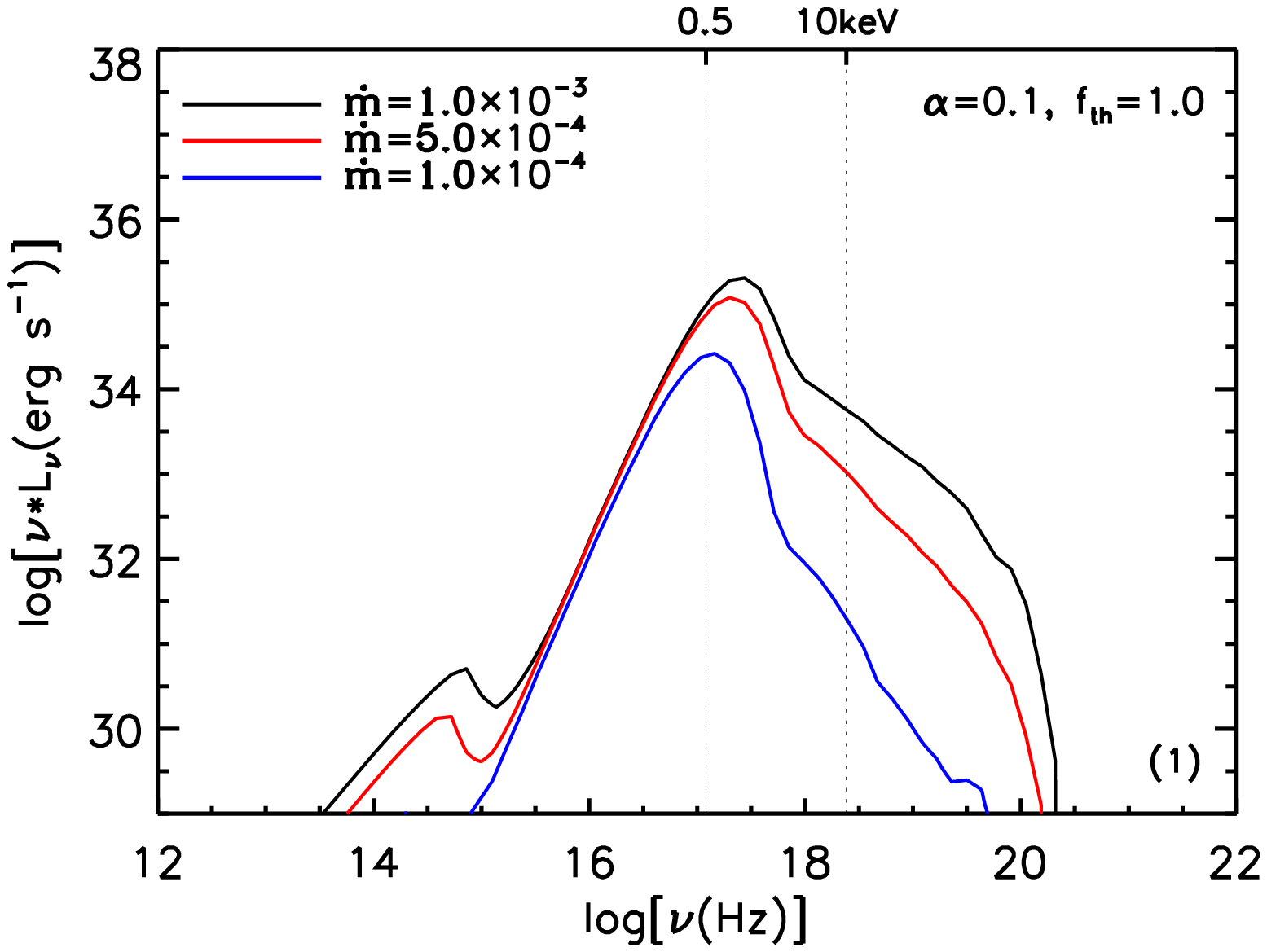}
\includegraphics[width=85mm,height=60mm,angle=0.0]{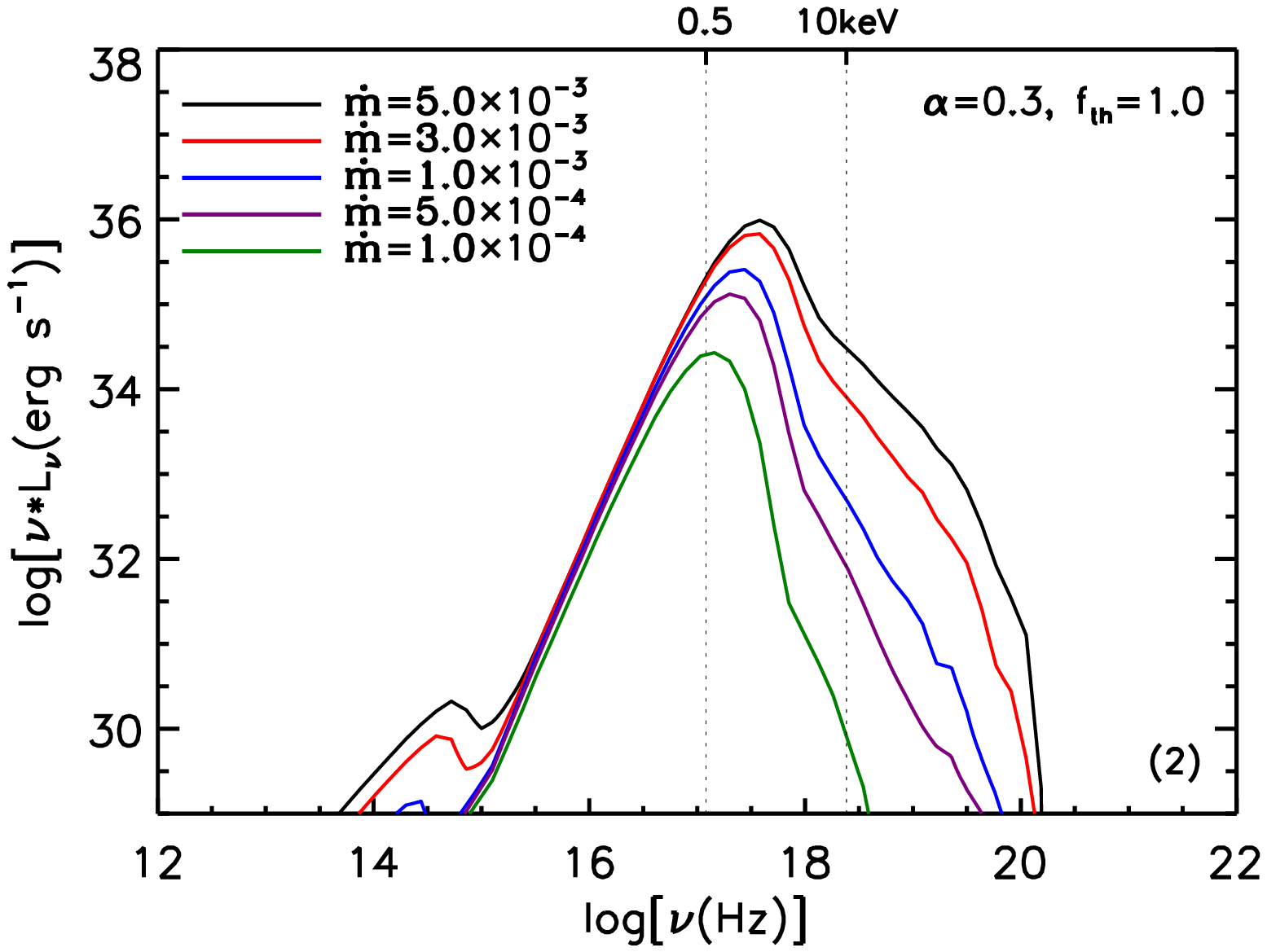}
\includegraphics[width=85mm,height=60mm,angle=0.0]{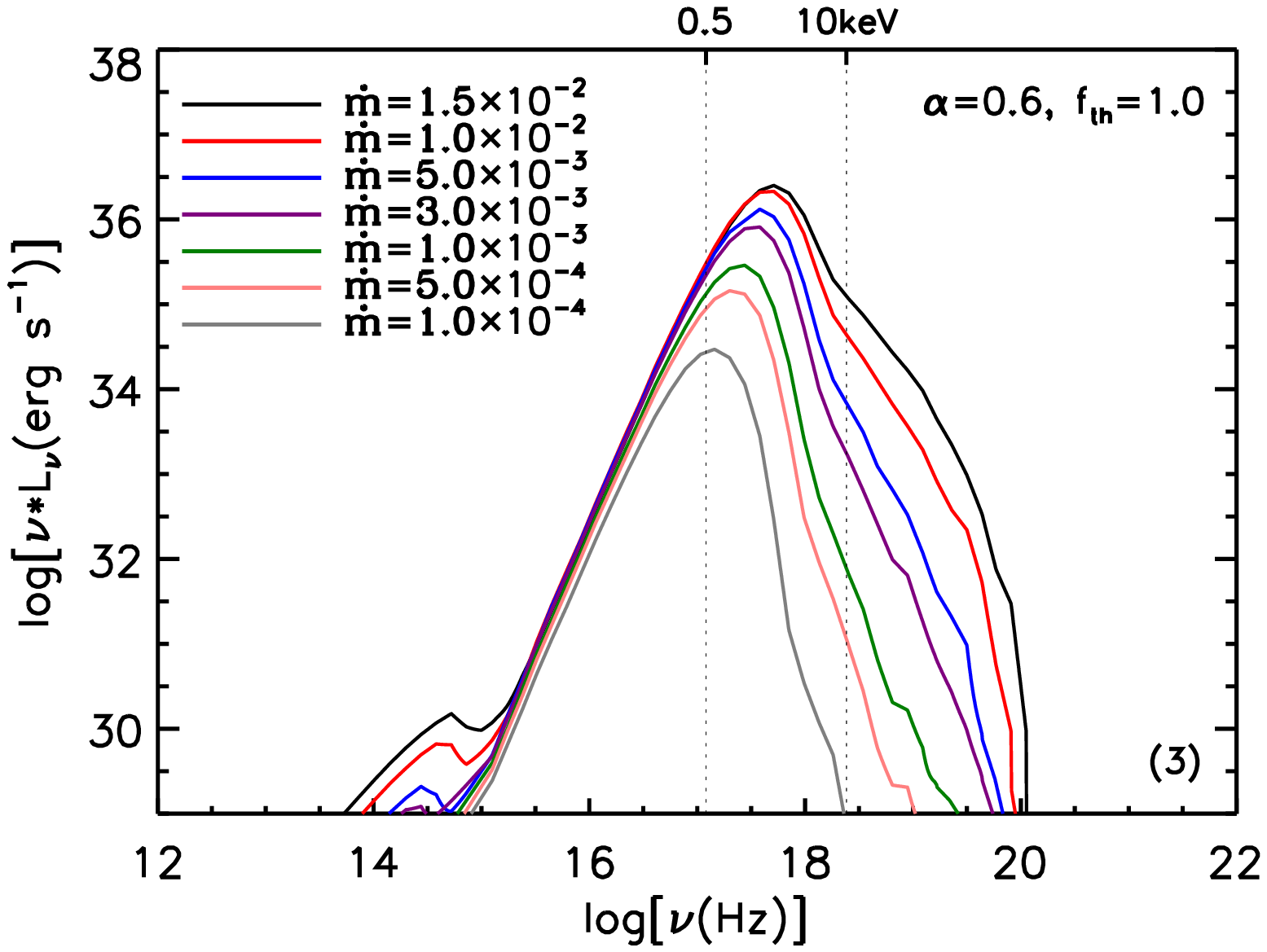}
\includegraphics[width=85mm,height=60mm,angle=0.0]{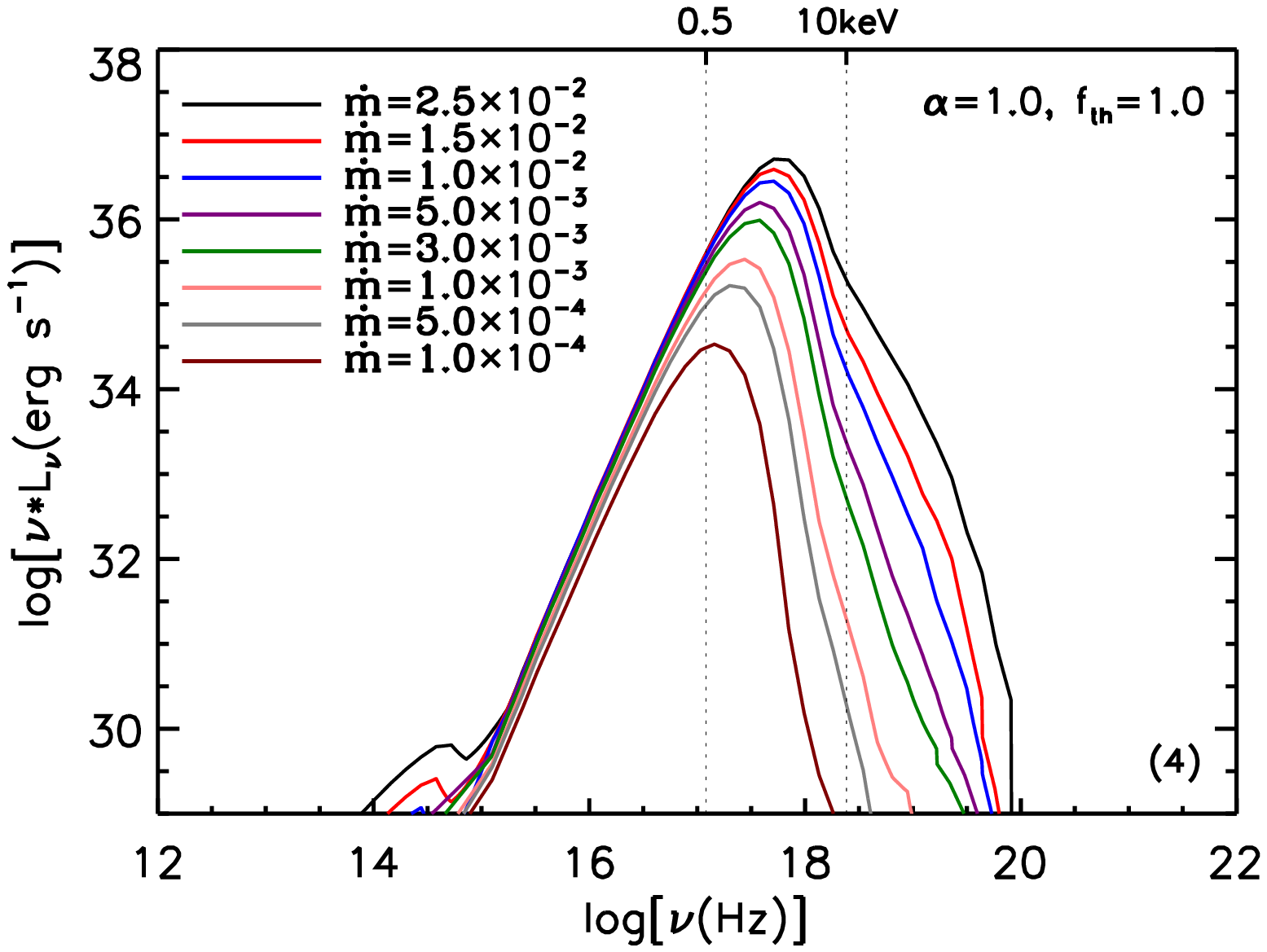}
\caption{\label{f:sp_alpha}Panel (1): Emergent spectra of the ADAF around a NS for $\alpha=0.1$ and $f_{\rm th}=1.0$.  
Panel (2): Emergent spectra of the ADAF around a NS for $\alpha=0.3$ and $f_{\rm th}=1.0$.
Panel (3): Emergent spectra of the ADAF around a NS for $\alpha=0.6$ and $f_{\rm th}=1.0$.
Panel (4): Emergent spectra of the ADAF around a NS for $\alpha=1.0$ and $f_{\rm th}=1.0$.
}
\end{figure*}

\begin{figure}
\includegraphics[width=85mm,height=60mm,angle=0.0]{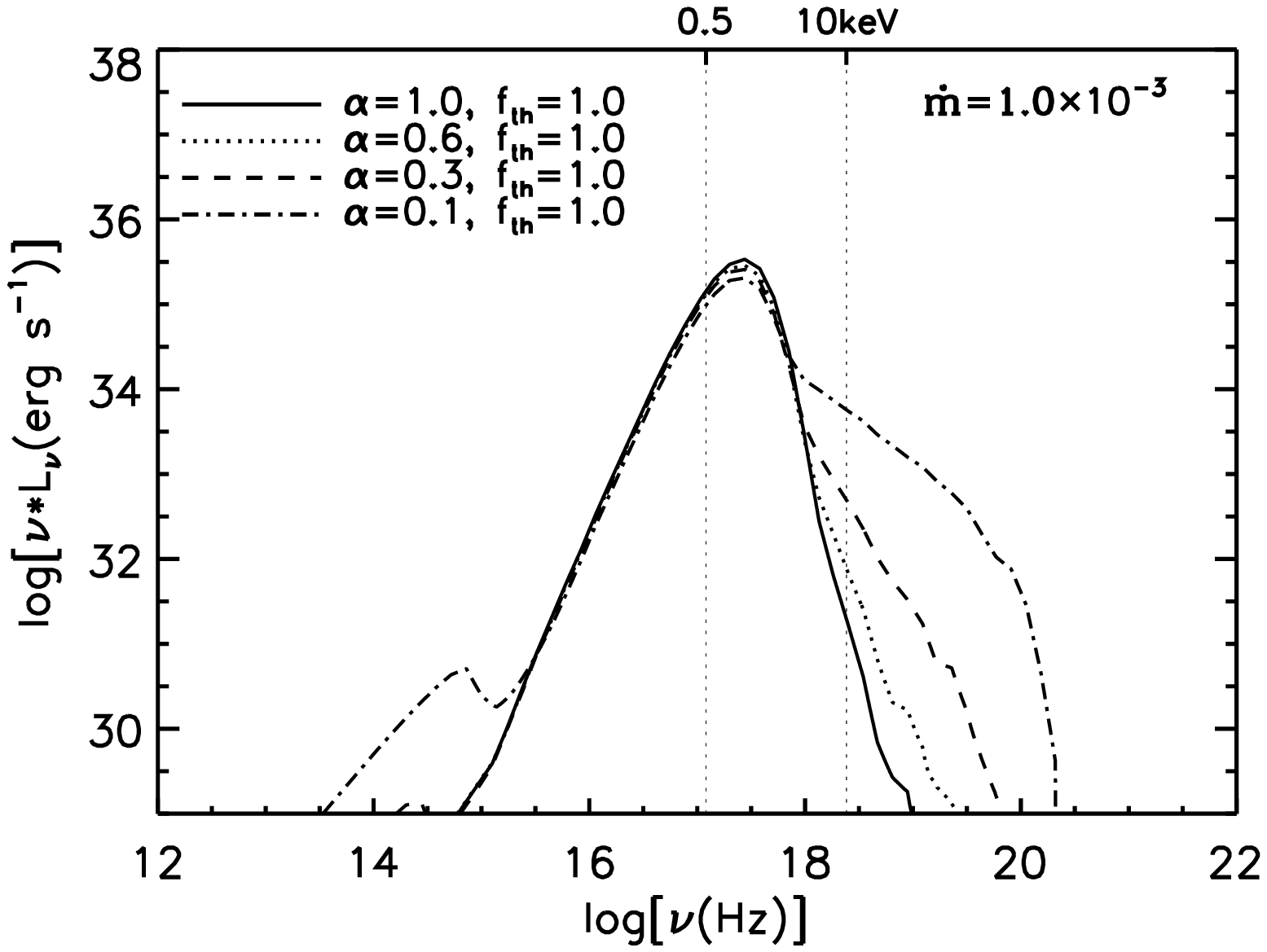}
\caption{\label{f:sp_difalpha}Emergent spectra of the ADAF around a NS for different $\alpha$ with
$f_{\rm th}=1.0$ and $\dot m=1\times 10^{-3}$. }
\end{figure}

In order to more easily compare with observations, in the panel (3) of 
Fig. \ref{f:alpha}, we plot the corresponding fractional contribution of the power-law luminosity 
$\eta$ as a function of the X-ray luminosity $L_{\rm 0.5-10\rm keV}$ for different $\alpha$ with $f_{\rm th}=1.0$. 
It can be seen that $\eta$ is nearly a constant ($\sim$ zero) with $L_{\rm 0.5-10\rm keV}$ for all the value of $\alpha$ as 
taking $\alpha=0.1, 0.3, 0.6$ and $1.0$ respectively, which means that the X-ray spectrum in the range 
of $0.5-10$ keV is completely dominated by the thermal component from the surface of the NS, and the 
contribution of the power-law component from the ADAF itself to the total X-ray luminosity 
$L_{\rm 0.5-10\rm keV}$ can nearly be neglected. 
One can refer to Table \ref{T:alpha_effect} for the detailed numerical results.
Meanwhile, one can refer to the detailed emergent spectra 
in the panel (1) of Fig. \ref {f:sp_alpha} for $\alpha=0.1$ and $f_{\rm th}=1.0$,   
in the panel (2) of Fig. \ref {f:sp_alpha} for $\alpha=0.3$ and $f_{\rm th}=1.0$, 
in the panel (3) of Fig. \ref {f:sp_alpha} for $\alpha=0.6$ and $f_{\rm th}=1.0$, and 
in the panel (4) of Fig. \ref {f:sp_alpha} for $\alpha=1.0$ and $f_{\rm th}=1.0$. 
In order to more clearly show the effect of $\alpha$ on the emergent spectra, in 
Fig. \ref {f:sp_difalpha}, we plot the emergent spectra for different $\alpha$ with
$\dot m=1\times 10^{-3}$ and $f_{\rm th}=1.0$. As we expect, the relative contribution of the  
power-law luminosity from the ADAF itself to the thermal luminosity from the surface of the NS decreases 
with increasing with $\alpha$. However, we should also keep in mind that the X-ray spectrum in the range of 
$0.5-10$ keV is completely dominated by the thermal component from the surface of the NS for all the value 
of $\alpha$ as taking $\alpha=0.1, 0.3, 0.6$ and $1.0$ respectively.
In the panel (4) of Fig. \ref{f:alpha}, we plot the corresponding temperature at the surface of the 
NS $T_{*}$ as a function of the X-ray luminosity $L_{\rm 0.5-10\rm keV}$ for different $\alpha$ with 
$f_{\rm th}=1.0$. It is clear that $T_{*}$ decreases with decreasing $L_{\rm 0.5-10\rm keV}$ for 
$\alpha=0.1, 0.3, 0.6$ and $1.0$ respectively. One can also refer to Table \ref{T:alpha_effect} for 
the detailed numerical results.

\begin{table*}
\caption{Radiative features of the ADAF around NSs for different $\dot m$ with 
$\alpha=0.1, 0.3, 0.6$ and $1.0$ respectively. 
$L_{*}/L_{\rm G}$ is the ratio of the energy of the ADAF transferred onto the 
surface of the NS per second to the accretion luminosity. 
$T_{*}$ is the effective temperature at the surface of the NS.
$L^{\rm power\ law}_{\rm 0.5-10\rm keV}/L_{\rm 0.5-10\rm keV}$ is the fractional 
contribution of the power-law luminosity.
$L_{\rm 0.5-10 keV}$ is the luminosity between 0.5 and 10 $\rm keV$.}
\centering
\begin{tabular}{ccccccc}
\hline\hline
\multicolumn{6}{l}{$m=1.4$, $R_{*}=12.5\ \rm km$, $\beta=0.95$ and $\nu_{\rm NS}=0\ (\rm Hz)$} \\
\hline
$\alpha$ & $f_{\rm th}$ & $\dot m$  & $L_{*}/L_{\rm G}$  & $T_{*} \ (\rm keV)$  &
$\eta$ (\%) & $L_{\rm 0.5-10 keV}\ (\rm erg \ s^{-1}) $ \\
\hline
0.1   & 1.0  &  $1.0\times10^{-3}$     &60.1\%    &0.31  &1.5 &  $2.5\times 10^{35}$ \\
0.1   & 1.0  &  $5.0\times10^{-4}$     &60.8\%    &0.26  &0.1 &  $1.3\times 10^{35}$ \\
0.1   & 1.0  &  $1.0\times10^{-4}$     &61.2\%    &0.17  &1.3 &  $1.9\times 10^{34}$ \\
\hline
0.3   & 1.0  &  $5.0\times10^{-3}$     &61.4\%    &0.46  &0.1 &  $1.3\times 10^{36}$ \\
0.3   & 1.0  &  $3.0\times10^{-3}$     &62.0\%    &0.40  &0.1 &  $8.7\times 10^{35}$ \\
0.3   & 1.0  &  $1.0\times10^{-3}$     &62.5\%    &0.31  &0.2 &  $3.0\times 10^{35}$ \\
0.3   & 1.0  &  $5.0\times10^{-4}$     &62.6\%    &0.26  &0.2 &  $1.4\times 10^{35}$ \\
0.3   & 1.0  &  $1.0\times10^{-4}$     &62.7\%    &0.17  &0.2 &  $1.9\times 10^{34}$ \\
\hline
0.6   & 1.0  &  $1.5\times10^{-2}$     &65.2\%    &0.61  &0.1 &  $3.5\times 10^{36}$ \\
0.6   & 1.0  &  $1.0\times10^{-2}$     &66.1\%    &0.56  &0.1 &  $3.0\times 10^{36}$ \\
0.6   & 1.0  &  $5.0\times10^{-3}$     &66.8\%    &0.47  &0.1 &  $1.6\times 10^{36}$ \\
0.6   & 1.0  &  $3.0\times10^{-3}$     &67.1\%    &0.41  &0.1 &  $1.0\times 10^{36}$ \\
0.6   & 1.0  &  $1.0\times10^{-3}$     &67.3\%    &0.31  &0.1 &  $3.3\times 10^{35}$ \\
0.6   & 1.0  &  $5.0\times10^{-4}$     &67.4\%    &0.26  &0.1 &  $1.5\times 10^{35}$ \\
0.6   & 1.0  &  $1.0\times10^{-4}$     &67.4\%    &0.18  &0.1 &  $2.1\times 10^{34}$ \\
\hline
1.0   & 1.0  &  $2.5\times10^{-2}$     &75.5\%    &0.72  &0.1 &  $7.3\times 10^{36}$ \\
1.0   & 1.0  &  $1.5\times10^{-2}$     &76.8\%    &0.64  &0.2 &  $5.2\times 10^{36}$ \\
1.0   & 1.0  &  $1.0\times10^{-2}$     &77.4\%    &0.58  &0.2 &  $3.8\times 10^{36}$ \\
1.0   & 1.0  &  $5.0\times10^{-3}$     &78.0\%    &0.49  &0.2 &  $2.0\times 10^{36}$ \\
1.0   & 1.0  &  $3.0\times10^{-3}$     &78.2\%    &0.43  &0.1 &  $1.2\times 10^{36}$ \\
1.0   & 1.0  &  $1.0\times10^{-3}$     &78.4\%    &0.33  &0.1 &  $3.9\times 10^{35}$ \\
1.0   & 1.0  &  $5.0\times10^{-4}$     &78.5\%    &0.27  &0.1 &  $1.8\times 10^{35}$ \\
1.0   & 1.0  &  $1.0\times10^{-4}$     &78.5\%    &0.18  &0.1 &  $2.6\times 10^{34}$ \\
\hline\hline
\end{tabular}
\\
\label{T:alpha_effect}
\end{table*}

\subsection{The effect of $f_{\rm th}$}\label{Sec:fth}
\begin{figure*}
\includegraphics[width=85mm,height=60mm,angle=0.0]{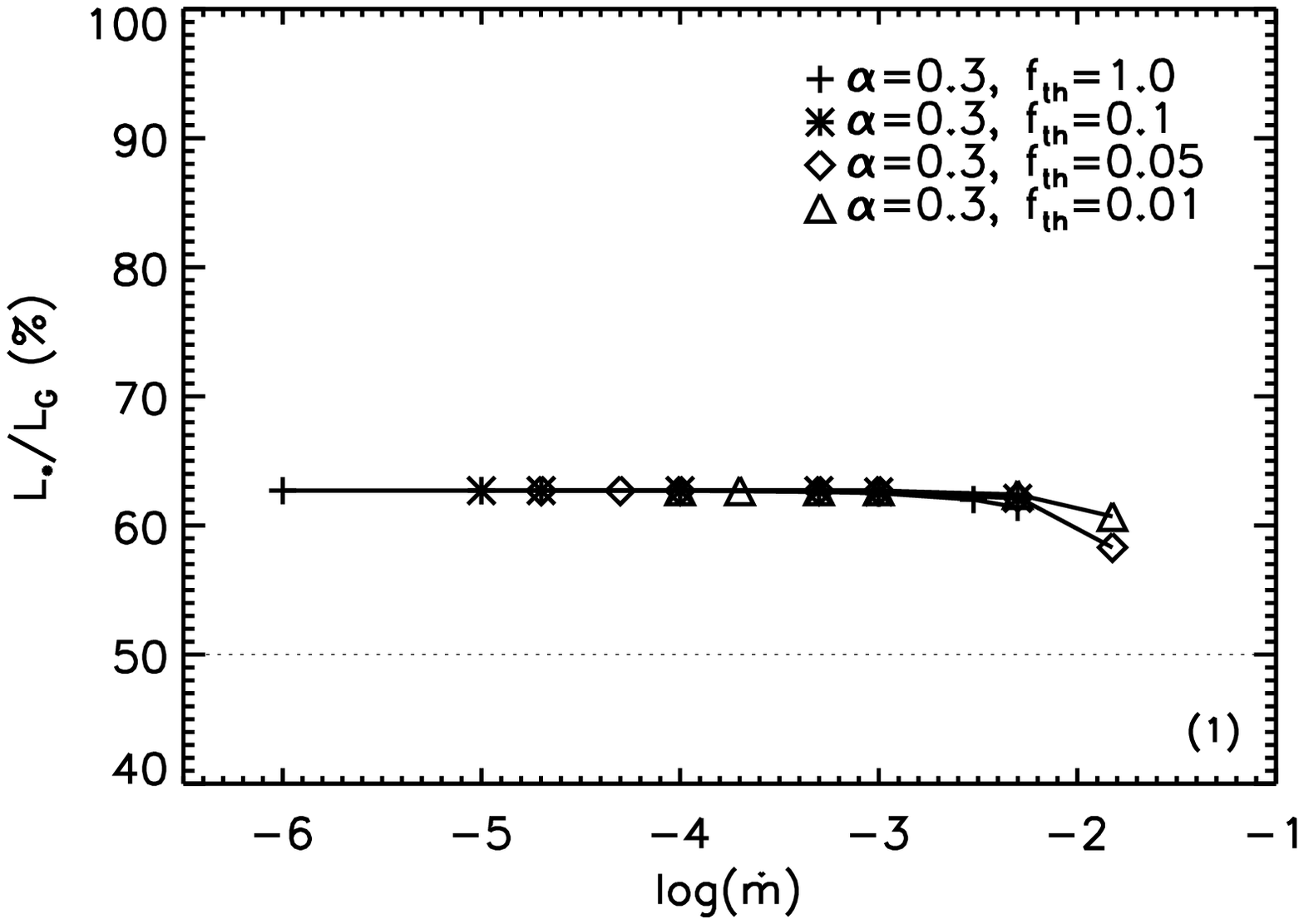}
\includegraphics[width=85mm,height=60mm,angle=0.0]{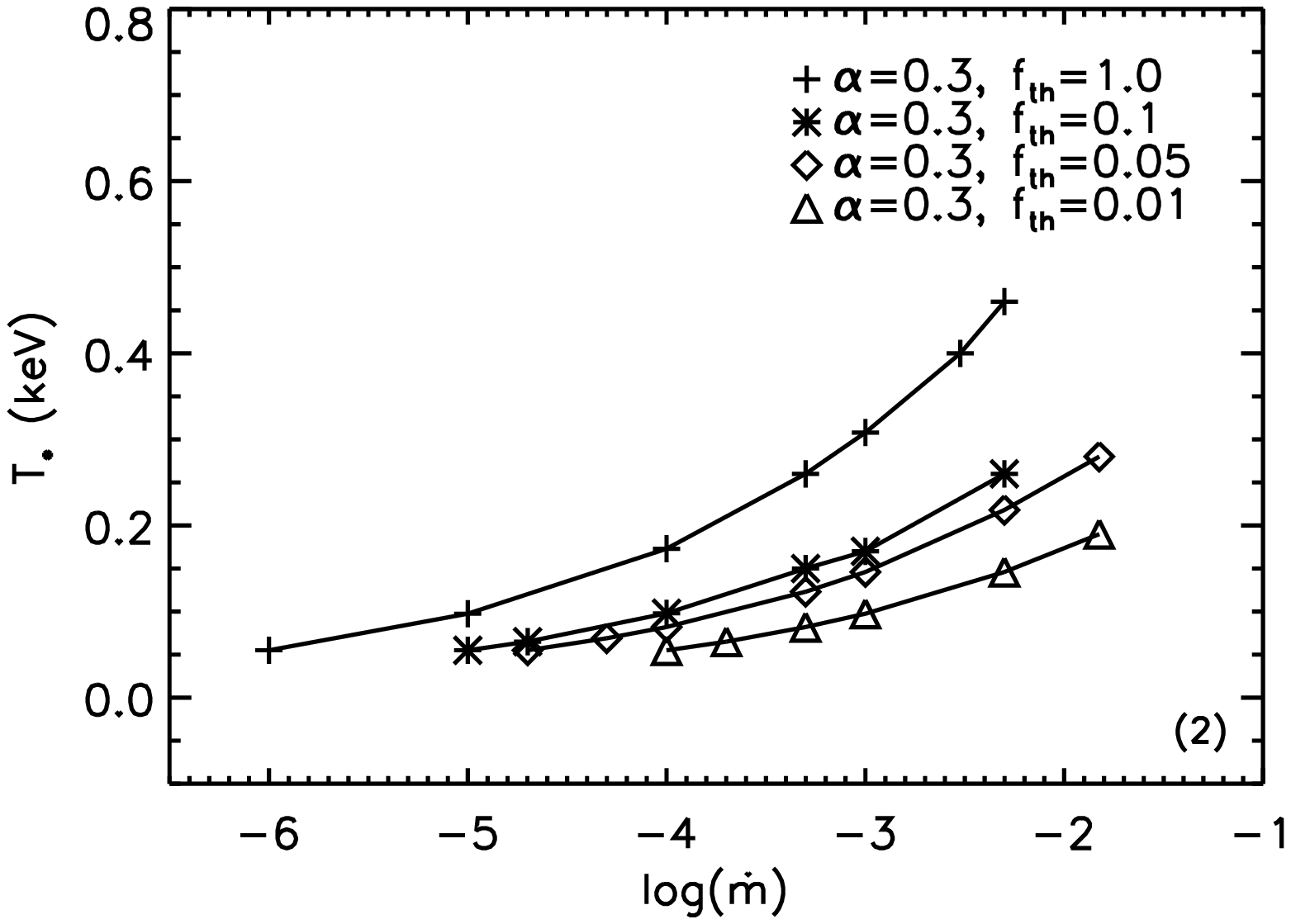}
\includegraphics[width=85mm,height=60mm,angle=0.0]{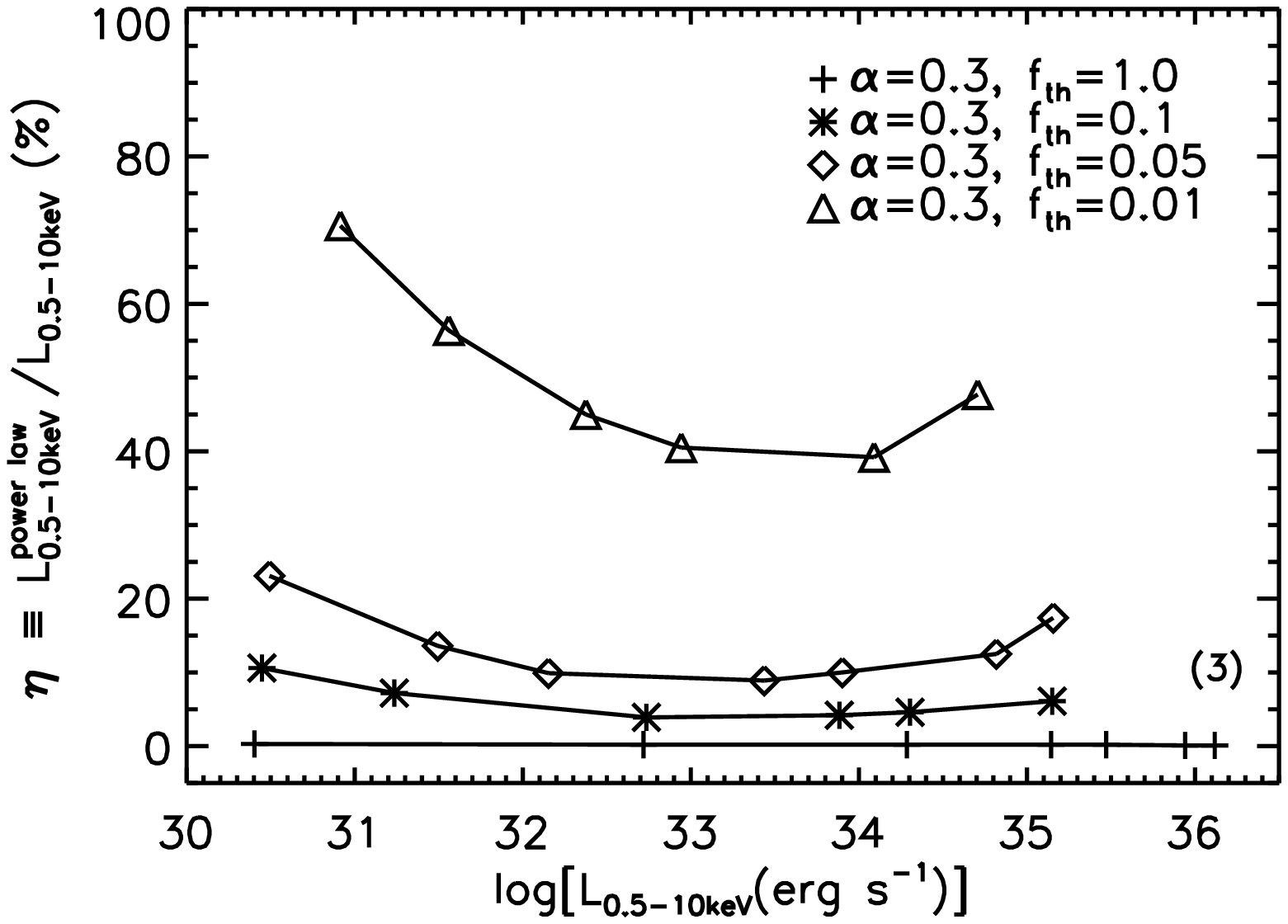}
\includegraphics[width=85mm,height=60mm,angle=0.0]{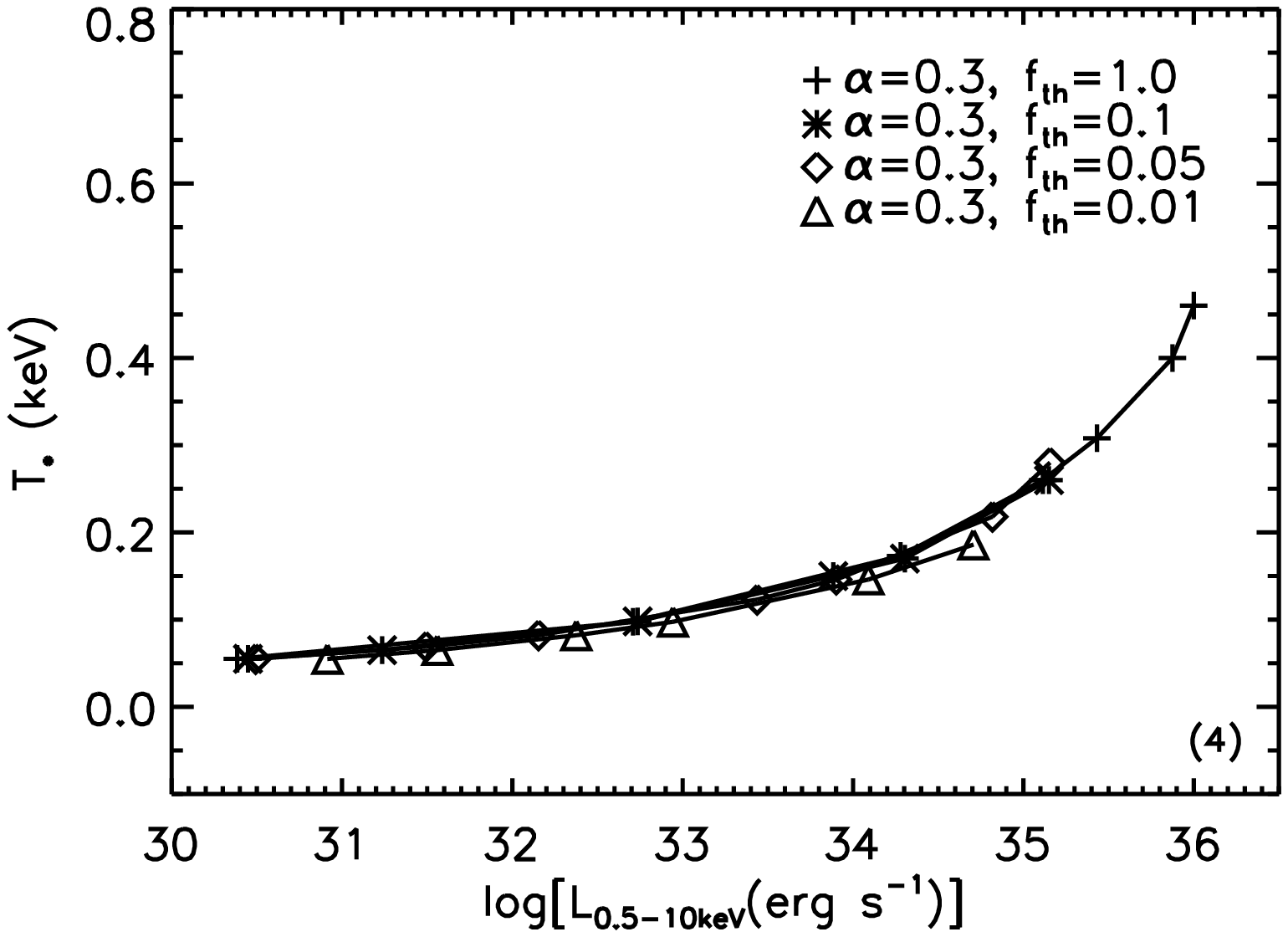}
\caption{\label{f:fth}Panel (1): Ratio of the energy transferred onto
the surface of the NS per second $L_{*}$ to the accretion luminosity $L_{\rm G}$, 
$L_{*}/L_{\rm G}$, as a function of $\dot m$.
Panel (2): Effective temperature at the surface of the NS $T_{*}$ as a function of $\dot m$.  
Panel (3): Fractional contribution of the power-law luminosity 
$\eta\equiv L^{\rm power\ law}_{\rm 0.5-10\rm keV}/L_{\rm 0.5-10\rm keV}$ as a function of the X-ray luminosity  
$L_{\rm 0.5-10\rm keV}$.
Panel (4): Effective temperature at the surface of the NS $T_{*}$ as a function of the 
X-ray luminosity  $L_{\rm 0.5-10\rm keV}$.
 }
\end{figure*}

In the panel (1) of Fig. \ref {f:fth},  we plot $L_{*}/L_{\rm G}$ as a function of $\dot m$ for
different $f_{\rm th}$ with $\alpha=0.3$. 
It can be seen that $L_{*}/L_{\rm G}$ increases slightly  with decreasing  $\dot m$ 
for taking $f_{\rm th}=1.0, 0.1, 0.05$ and $0.01$ respectively. Meanwhile, as we can see that  
the effect of $f_{\rm th}$ on the relation between  $L_{*}/L_{\rm G}$ and $\dot m$ is very little. 
One can refer to Table \ref{T:fth_effect} for the detailed numerical results. 
In the panel (2) of Fig. \ref {f:fth},  we plot $T_{*}$ as a function of $\dot m$
for different $f_{\rm th}$ with $\alpha=0.3$. 
It can be seen that $T_{*}$ decreases with decreasing $\dot m$ for $f_{\rm th}=1.0, 0.1, 0.05$ 
and $0.01$ respectively.
Meanwhile, it is clear that, the value of $T_{*}$ for a bigger $f_{\rm th}$  is 
systematically higher than that of for a smaller value of $f_{\rm th}$ for a 
fixed $\dot m$,  as can be easily derived from equation \ref {equ:T} in this paper.

\begin{figure*}
\includegraphics[width=85mm,height=60mm,angle=0.0]{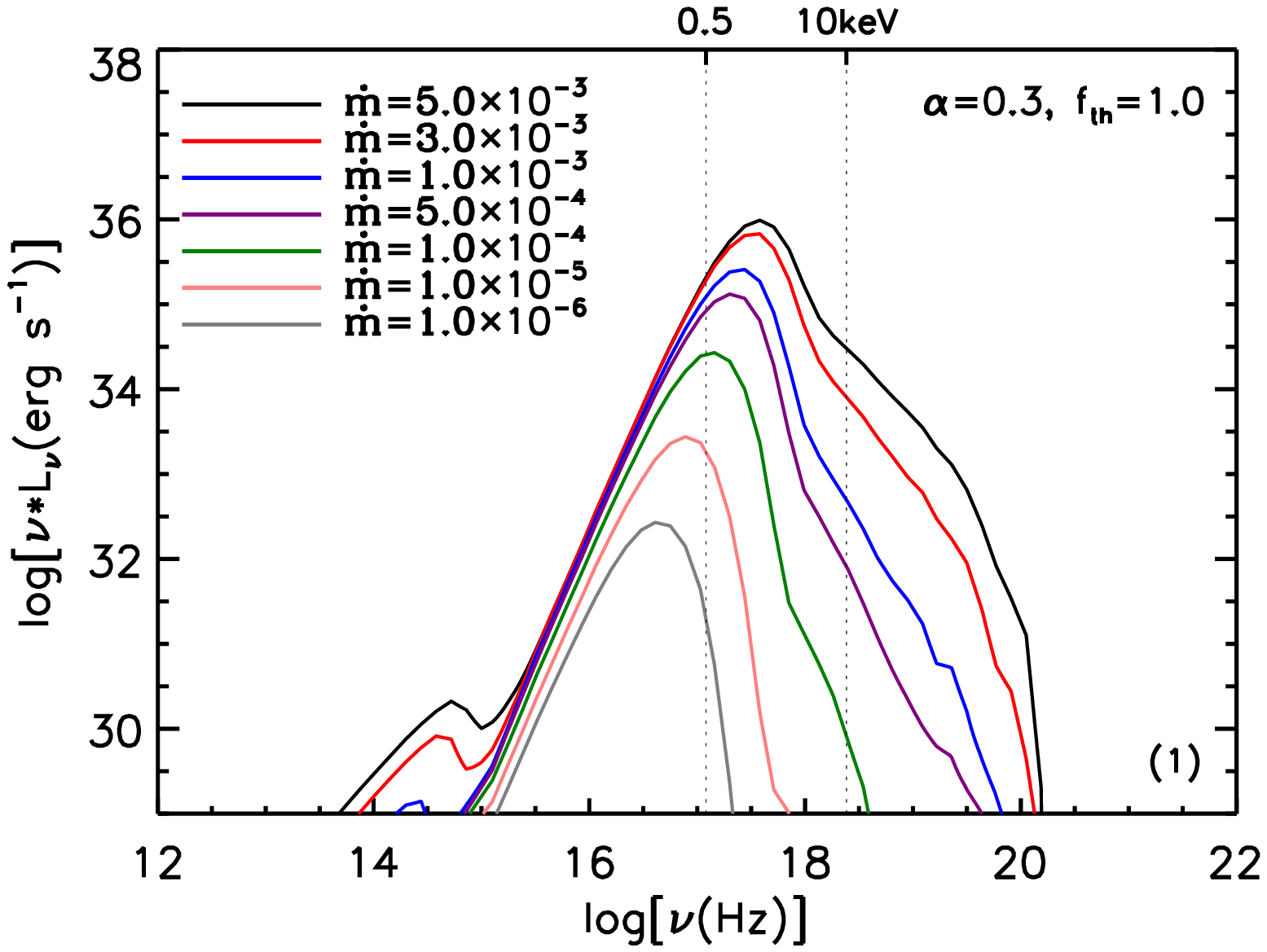}
\includegraphics[width=85mm,height=60mm,angle=0.0]{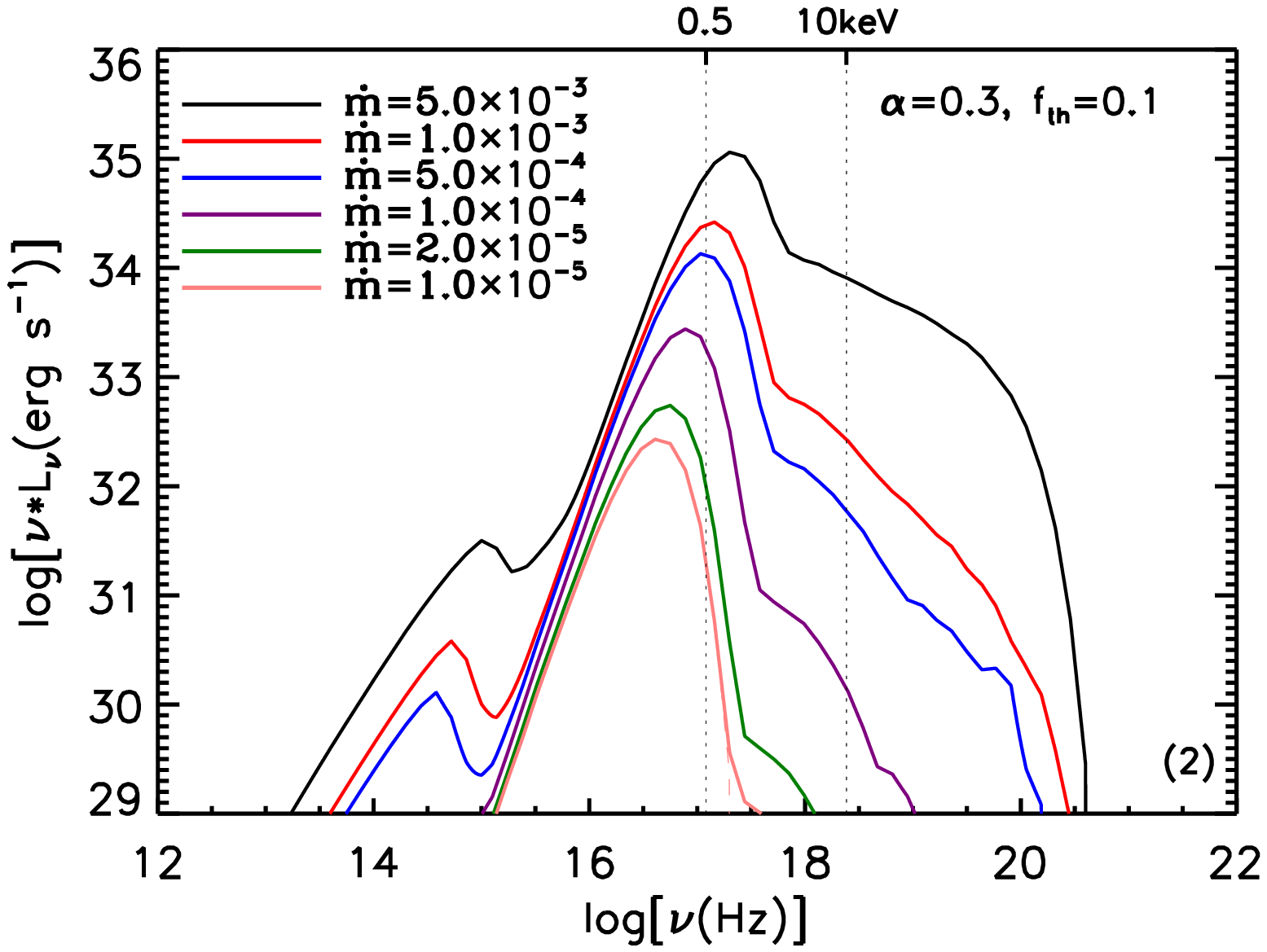}
\includegraphics[width=85mm,height=60mm,angle=0.0]{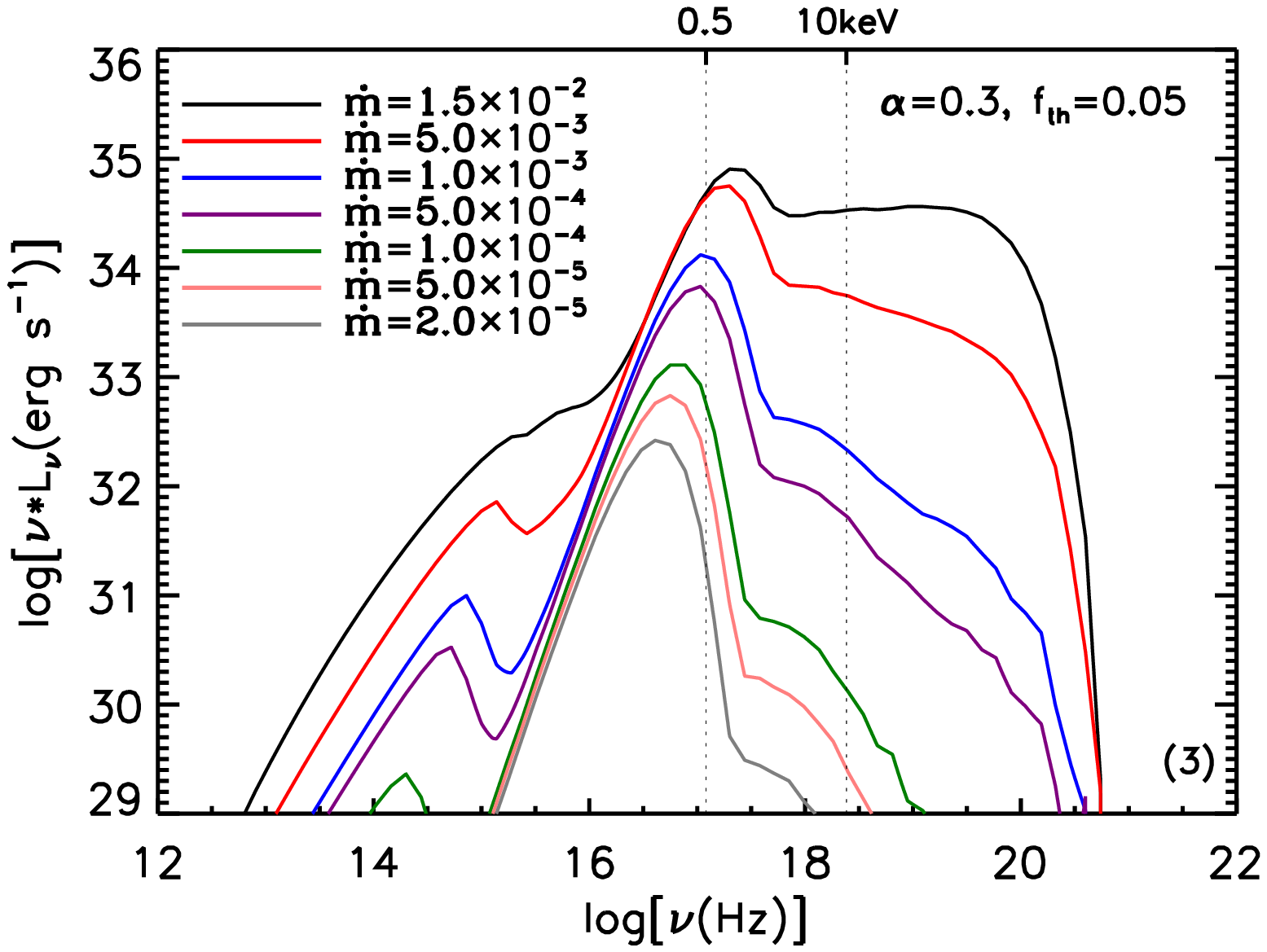}
\includegraphics[width=85mm,height=60mm,angle=0.0]{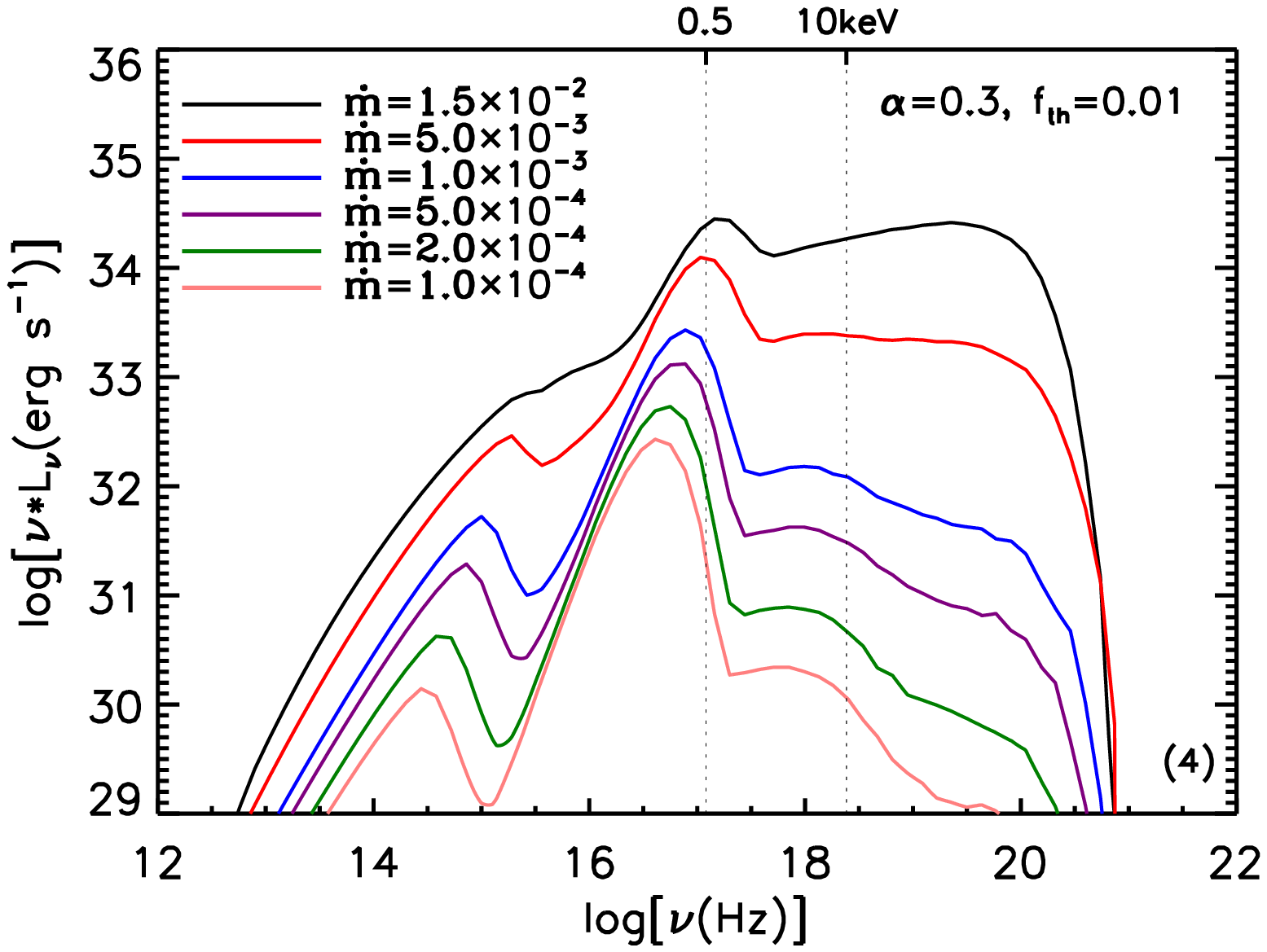}
\caption{\label{f:sp_fth}Panel (1): Emergent spectra of the ADAF around a NS for $\alpha=0.3$ and $f_{\rm th}=1.0$.
Panel (2): Emergent spectra of the ADAF around a NS for $\alpha=0.3$ and $f_{\rm th}=0.1$.
Panel (3): Emergent spectra of the ADAF around a NS for $\alpha=0.3$ and $f_{\rm th}=0.05$.
Panel (4): Emergent spectra of the ADAF around a NS for $\alpha=0.3$ and $f_{\rm th}=0.01$.
}
\end{figure*}

\begin{figure}
\includegraphics[width=85mm,height=60mm,angle=0.0]{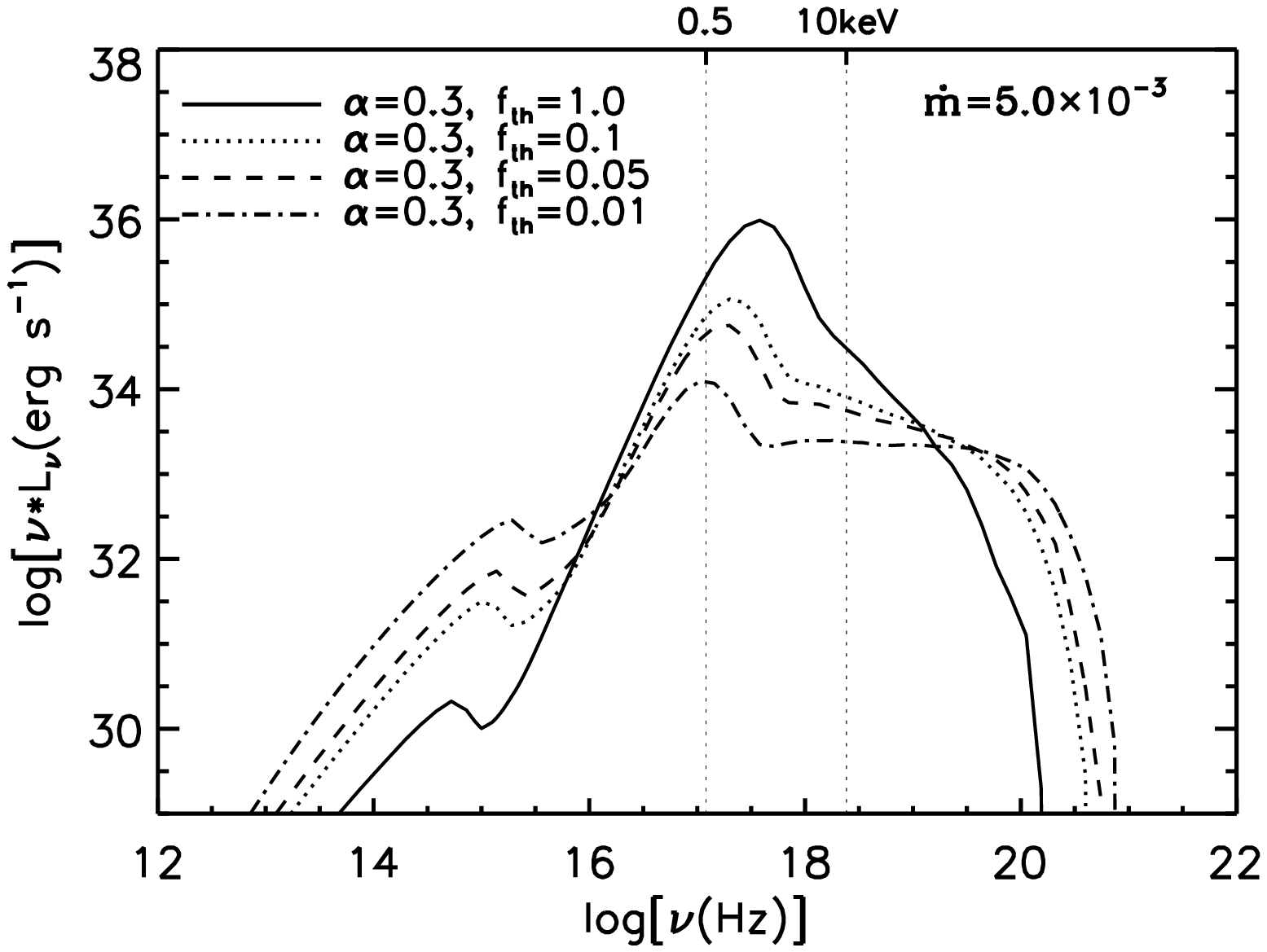}
\caption{\label{f:sp_diffth} Emergent spectra of the ADAF around a NS for different $f_{\rm th}$ with
$\alpha=0.3$ and $\dot m=5\times 10^{-3}$.  }
\end{figure}

In order to more easily compare with observations, in the panel (3) of Fig. \ref {f:fth}, 
we plot the corresponding fractional contribution of the power-law luminosity $\eta$ as a function 
of the X-ray luminosity $L_{\rm 0.5-10\rm keV}$ for different $f_{\rm th}$ with $\alpha=0.3$.  
It can be seen that, for $f_{\rm th}=0.1$, 
$\eta$ decreases from $6.1\%$ to $3.9\%$ for $L_{\rm 0.5-10\rm keV}$ decreasing from
$1.4\times 10^{35}\ \rm erg \ s^{-1}$ to $5.4\times 10^{32}\ \rm erg \ s^{-1}$,
then $\eta$ increases from $3.9\%$ to $10.6\%$ for $L_{\rm 0.5-10\rm keV}$ decreasing from
$5.4\times 10^{32}\ \rm erg \ s^{-1}$ to $2.8\times 10^{30}\ \rm erg \ s^{-1}$
\footnote{In this paper, we always describe the relation between $\eta$ and 
$L_{\rm 0.5-10\rm keV}$ from a higher value of $L_{\rm 0.5-10\rm keV}$ to a lower value of 
$L_{\rm 0.5-10\rm keV}$. This is because the quiescent spectrum is often observed after the 
outburst as the order that the X-ray luminosity decreases with time.}.
A similar trend between $\eta$ and $L_{\rm 0.5-10\rm keV}$ is shown for $f_{\rm th}=0.05$ and
$f_{\rm th}=0.01$ respectively as for $f_{\rm th}=0.1$ (As a comparison, 
$\eta$ versus $L_{\rm 0.5-10\rm keV}$ is also plotted for $f_{\rm th}=1.0$). 
One can refer to Table \ref{T:fth_effect} for the detailed numerical results.
Meanwhile, one can refer to the detailed emergent spectra
in the panel (1) of Fig. \ref{f:sp_fth} for $\alpha=0.3$ and $f_{\rm th}=1.0$, 
in the panel (2) of Fig. \ref{f:sp_fth} for $\alpha=0.3$ and $f_{\rm th}=0.1$, 
in the panel (3) of Fig. \ref{f:sp_fth} for $\alpha=0.3$ and $f_{\rm th}=0.05$, and 
in the panel (4) of Fig. \ref{f:sp_fth} for $\alpha=0.3$ and $f_{\rm th}=0.01$.
In order to more clearly show the effect of $f_{\rm th}$ on the emergent spectra, we plot the 
emergent spectra for different $f_{\rm th}$ with $\dot m=5\times 10^{-3}$ and $\alpha=0.3$ 
in Fig. \ref {f:sp_diffth}. 
As we expect, both the temperature and luminosity of
the thermal component decrease with decreasing $f_{\rm th}$.  
As we can see from Fig. \ref {f:sp_diffth}, the luminosity of the power-law component
decreases with decreasing $f_{\rm th}$, and the slope of the power-law component also decreases with 
decreasing $f_{\rm th}$, which can be roughly understood as follows.
As we know, in the NS case, the Compton cooling of the ADAF is dominated by the cooling of the  
seed photons from the surface of the NS, and the Compton cooling rate can be roughly expressed as, 
$q_{\rm Cmp}={4kT_{\rm e}\over {m_{\rm e}c^2}}n_{\rm e}\sigma_{T}cu \propto ycu/H$
(with $k$ being the Boltzmann constant,  $T_{\rm e}$ being  the electron temperature, 
$m_{\rm e}$ being the electron mass, $c$ being the speed of light, $n_{\rm e}$ being the 
electron number density, $\sigma_{\rm T}$ being the Thomson cross section,  
$H$ being the scaleheight of the ADAF, $y$ being the
Compton $y$-parameter, and $u$ being the energy density of the seed photons from the surface of the NS).
With a decrease of $f_{\rm th}$ from $f_{\rm th}=1.0$ to $f_{\rm th}=0.01$, 
the seed photon energy density $u$ will decrease by a factor of $\sim 100$, and $y$ will increase  
by a factor of only a few times [see the panel (3) of Fig. 5 in \citet[][]{Qiao2018b}]. 
Finally, there will be a few tens of the decrease of the power-law 
luminosity. As aforementioned, with a decrease of $f_{\rm th}$, $y$ will increase, which will lead to
a harder X-ray spectrum, i.e., the slope of the power-law component will decrease with decreasing $f_{\rm th}$. 
In the panel (4) of Fig. \ref{f:fth}, we plot the corresponding temperature at the surface of the 
NS $T_{*}$ as a function of the X-ray luminosity $L_{\rm 0.5-10\rm keV}$ for different $f_{\rm th}$ with $\alpha=0.3$. 
It is clear that $T_{*}$ decreases with decreasing $L_{\rm 0.5-10\rm keV}$ for $f_{\rm th}=1.0, 0.1, 0.05$ and $0.01$ 
respectively, the trend of which is roughly consistent with observations.
One can also refer to Table \ref{T:fth_effect} for the detailed numerical results.

Here, we would like to argue that the effects of $f_{\rm th}$ 
on the relation between $\eta$ and $L_{\rm 0.5-10\rm keV}$ can be understood as follows.
As we can see from Fig. \ref {f:sp_diffth}, for a fixed $\dot m$, although the power-law luminosity from 
the ADAF decreases with decreasing  $f_{\rm th}$, the thermal luminosity from the surface of the NS
decreases more quickly, which actually means that the ratio of the power-law luminosity to the thermal 
luminosity increases with decreasing $f_{\rm th}$. Specifically, if we focus on the energy range between 
$0.5$ and $10$ keV, for $f_{\rm th}=0.01, 0.05, 0.1$ and $1.0$, $\eta$ is $39.2\%$, $12.5\%$, $6.1\%$ 
and  $0.1\%$ respectively for $\dot m=5\times 10^{-3}$.
For $f_{\rm th}=0.01$, if we increase the mass accretion rate as  
$\dot m > 5\times 10^{-3}$, the ratio of the  power-law luminosity to the thermal 
luminosity increases, and the value of $\eta$ increases with increasing $L_{\rm 0.5-10\rm keV}$,
i.e., there is a positive correlation between $\eta$ and $L_{\rm 0.5-10\rm keV}$. 
However, we note that, for $\dot m < 1\times 10^{-3}$
corresponding $L_{\rm 0.5-10\rm keV} < 8.8\times 10^{32}\ \rm erg \ s^{-1}$, 
although the ratio of the  power-law luminosity to the thermal luminosity decreases with decreasing 
$\dot m$, if we focus on the energy range between $0.5$ and $10$ keV, $\eta$ increases with 
decreasing $L_{\rm 0.5-10\rm keV}$, i.e., there is an anti-correlation between $\eta$ 
and $L_{\rm 0.5-10\rm keV}$. This is because, the temperature of the thermal component at the surface 
of the NS $T_{*}$ decreases with decreasing $\dot m$, especially, for $\dot m < 1\times 10^{-3}$, 
the peak emission of the thermal component from the surface of the NS moves out the 
range of $0.5-10\rm keV$. So although the ratio of the power-law luminosity to the thermal 
luminosity decreases with decreasing $\dot m$, if we focus on the energy range between $0.5$ and $10$ keV, 
$\eta$ increases with decreasing $\dot m$. Meanwhile, because $L_{\rm 0.5-10\rm keV}$ decreases with decreasing 
$\dot m$, $\eta$ increases with decreasing $L_{\rm 0.5-10\rm keV}$.
For example, for $\dot m=1\times 10^{-3}$, $T_{*}$ is $0.098$ keV, the peak emission is 
at $T_{\rm max}\approx 2.82 T_{*}\approx 0.28$ keV (in $L_{\rm \nu}$ versus $\nu$), a very minor fraction 
of the thermal component from the surface of the NS falls into the range of $0.5-10\rm keV$. The 
corresponding $\eta$ is $40.5\%$ and $L_{\rm 0.5-10\rm keV}$ is $8.8\times 10^{32}\ \rm erg \ s^{-1}$.
With the decrease of $\dot m$ further, it is clear that $\eta$ increases with decreasing 
$L_{\rm 0.5-10\rm keV}$.

As we can see from the panel (3) of Fig. \ref {f:fth}, $\eta$ as a function of $L_{\rm 0.5-10\rm keV}$ for 
a bigger value of $f_{\rm th}=0.1$ is  systematically lower than that of for  $f_{\rm th}=0.05$ and $f_{\rm th}=0.01$.
Meanwhile, $f_{\rm th}=0.1$, there still exists a correlation between $\eta$ and $L_{\rm 0.5-10\rm keV}$, i.e.,
there is a positive correlation between $\eta$ and $L_{\rm 0.5-10\rm keV}$ for 
$L_{\rm 0.5-10\rm keV} \gtrsim$ a few times of $10^{33}\ \rm erg \ s^{-1}$, below which 
is an anti-correlation between $\eta$ and $L_{\rm 0.5-10\rm keV}$. 
Compared with the case for $f_{\rm th}=0.05$ and $f_{\rm th}=0.01$, the slope of both the positive correlation and the 
anti-correlation between $\eta$ and $L_{\rm 0.5-10\rm keV}$ becomes flatter, indicating that the correlation between 
$\eta$ and $L_{\rm 0.5-10\rm keV}$ becomes  weaker. Further, if we take the maximum value of 
$f_{\rm th}$ as  $f_{\rm th}=1.0$, $\eta$ is nearly a constant ($\sim$ zero ) with $L_{\rm 0.5-10\rm keV}$, 
and there is no correlation between $\eta$ and $L_{\rm 0.5-10\rm keV}$.

Finally, we would like to mention that the relation of $\eta$ as a function of $L_{\rm 0.5-10\rm keV}$ 
for $f_{\rm th}=1.0, 0.1, 0.05$ and $0.01$ with $\alpha=0.3$ will systematically shift rightwards for taking  
a bigger value of $\alpha$, and will systematically shift leftwards for taking a smaller value of $\alpha$ 
respectively. This is because an increase (or a decrease) of $\alpha$ will only lead to an increase (or a decrease) 
of the upper limit of the X-ray luminosity but will not systematically change the trend of the shape of the X-ray 
spectrum with the X-ray luminosity.

\begin{table*}
\caption{Radiative features of the ADAF around NSs for different $\dot m$ with $f_{\rm th}=1.0,0.1, 0.05$ and $0.01$ 
respectively. $L_{*}/L_{\rm G}$ is the ratio of the energy of the ADAF transferred onto the 
surface of the NS per second to the accretion luminosity. 
$T_{*}$ is the effective temperature at the surface of the NS.
$L^{\rm power\ law}_{\rm 0.5-10\rm keV}/L_{\rm 0.5-10\rm keV}$ is the fractional 
contribution of the power-law luminosity.
$L_{\rm 0.5-10 keV}$ is the luminosity between 0.5 and 10 $\rm keV$.}
\centering
\begin{tabular}{ccccccc}
\hline\hline
\multicolumn{6}{l}{$m=1.4$, $R_{*}=12.5\ \rm km$, $\beta=0.95$ and $\nu_{\rm NS}=0\ (\rm Hz)$} \\
\hline
$\alpha$ & $f_{\rm th}$ & $\dot m$  & $L_{*}/L_{\rm G}$  & $T_{*} \ (\rm keV)$  &
$\eta$ (\%) & $L_{\rm 0.5-10 keV}\ (\rm erg \ s^{-1}) $ \\
\hline
0.3   & 1.0  &  $5.0\times10^{-3}$     &61.4\%    &0.46   &0.1 &  $1.3\times 10^{36}$ \\
0.3   & 1.0  &  $3.0\times10^{-3}$     &62.0\%    &0.40   &0.1 &  $8.7\times 10^{35}$ \\
0.3   & 1.0  &  $1.0\times10^{-3}$     &62.5\%    &0.31   &0.2 &  $3.0\times 10^{35}$ \\
0.3   & 1.0  &  $5.0\times10^{-4}$     &62.6\%    &0.26   &0.2 &  $1.4\times 10^{35}$ \\
0.3   & 1.0  &  $1.0\times10^{-4}$     &62.7\%    &0.17   &0.2 &  $1.9\times 10^{34}$ \\
0.3   & 1.0  &  $1.0\times10^{-5}$     &62.7\%    &0.098   &0.2 &  $5.2\times 10^{32}$ \\
0.3   & 1.0  &  $1.0\times10^{-6}$     &62.7\%    &0.055  &0.3 &  $2.5\times 10^{30}$ \\
\hline
0.3   & 0.1  &  $5.0\times10^{-3}$     &62.1\%    &0.26    &6.1 &   $1.4\times 10^{35}$ \\
0.3   & 0.1  &  $1.0\times10^{-3}$     &62.6\%    &0.17    &4.5 &   $2.0\times 10^{34}$ \\
0.3   & 0.1  &  $5.0\times10^{-4}$     &62.7\%    &0.15    &4.2 &   $7.6\times 10^{33}$ \\
0.3   & 0.1  &  $1.0\times10^{-4}$     &62.7\%    &0.098   &3.9 &   $5.4\times 10^{32}$ \\
0.3   & 0.1  &  $2.0\times10^{-5}$     &62.7\%    &0.065   &7.2 &   $1.7\times 10^{31}$ \\
0.3   & 0.1  &  $1.0\times10^{-5}$     &62.7\%    &0.055   &10.6 &  $2.8\times 10^{30}$ \\
\hline
0.3   & 0.05  &  $1.5\times10^{-2}$     &58.3\%    &0.28   &17.4 &  $1.4\times 10^{35}$ \\
0.3   & 0.05  &  $5.0\times10^{-3}$     &62.2\%    &0.22   &12.5 &  $6.6\times 10^{34}$ \\
0.3   & 0.05  &  $1.0\times10^{-3}$     &62.7\%    &0.15   &10.0 &  $8.0\times 10^{33}$ \\
0.3   & 0.05  &  $5.0\times10^{-4}$     &62.7\%    &0.12   &8.90 &  $2.7\times 10^{33}$ \\
0.3   & 0.05  &  $1.0\times10^{-4}$     &62.7\%    &0.082  &9.90 &  $1.4\times 10^{32}$ \\
0.3   & 0.05  &  $5.0\times10^{-5}$     &62.7\%    &0.069  &13.6 &  $3.1\times 10^{31}$ \\
0.3   & 0.05  &  $2.0\times10^{-5}$     &62.7\%    &0.055  &23.1 &  $3.1\times 10^{30}$ \\
\hline
0.3   & 0.01  &  $1.5\times10^{-2}$     &60.7\%    &0.19   &47.7 &  $5.5\times 10^{34}$ \\
0.3   & 0.01  &  $5.0\times10^{-3}$     &62.4\%    &0.15   &39.2 &  $1.2\times 10^{34}$ \\
0.3   & 0.01  &  $1.0\times10^{-3}$     &62.7\%    &0.098  &40.5 &  $8.8\times 10^{32}$ \\
0.3   & 0.01  &  $5.0\times10^{-4}$     &62.7\%    &0.082  &45.0 &  $2.4\times 10^{32}$ \\
0.3   & 0.01  &  $2.0\times10^{-4}$     &62.7\%    &0.065  &56.4 &  $3.6\times 10^{31}$ \\
0.3   & 0.01  &  $1.0\times10^{-4}$     &62.7\%    &0.055  &71.6 &  $8.2\times 10^{30}$ \\
\hline\hline
\end{tabular}
\\
\label{T:fth_effect}
\end{table*}

\subsection{Comparison with observations for $\eta$ versus $L_{\rm 0.5-10\rm keV}$---the effect of $f_{\rm th}$ }\label{Sec:compare}
As has been shown in Section \ref{Sec:alpha}  and Section \ref{Sec:fth}, we test the 
effect of the viscosity parameter $\alpha$, and $f_{\rm th}$ describing the fraction 
of the total  energy of the ADAF transferred onto the surface of the NS to be thermalized as the thermal 
emission on the relation between $\eta$ and  $L_{\rm 0.5-10\rm keV}$ respectively. 
It is found that, theoretically, the effect of $\alpha$  on the relation between 
$\eta$ and  $L_{\rm 0.5-10\rm keV}$ is very little, and can nearly be neglected. Specifically, 
$\eta$ is nearly a constant ($\sim $ zero) with $L_{\rm 0.5-10\rm keV}$ for different $\alpha$ 
with $f_{\rm th}=1$, which is inconsistent with observations. 
So in this paper, we focus on the effect of $f_{\rm th}$ on the relation between $\eta$ and  
$L_{\rm 0.5-10\rm keV}$, and compare the theoretical results with observations.

\begin{figure}
\includegraphics[width=85mm,height=60mm,angle=0.0]{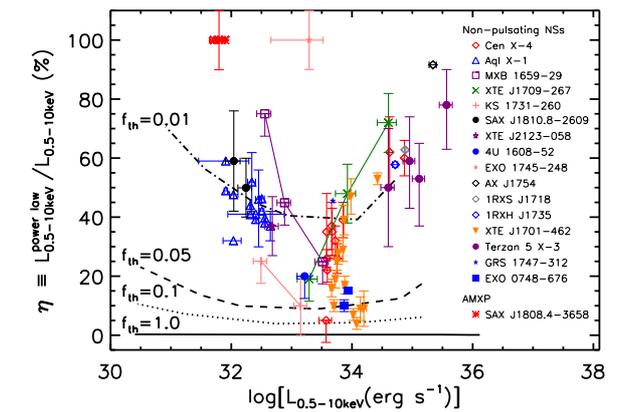}
\caption{\label{f:fth_obs} The fractional contribution of the power-law luminosity $\eta$ versus 
the X-ray luminosity $L_{\rm 0.5-10\rm keV}$. The solid line, the dotted line, the dashed line,
and the dot-dashed lines refer to the theoretical results for taking 
$f_{\rm th}=1.0, 0.1, 0.05$ and $0.01$ respectively. In the calculations, $\alpha=0.3$ is adopted.
The coloured symbols are the observational data of 16 non-pulsating NS-LMXBs probably dominated by
low-level accretion onto NSs. One AMXP SAX J1808.4-3658 is included for comparison.
One can refer to Section \ref{Sec:compare} for the references for details.  }
\end{figure}

We collect the observational data of $\eta$ versus $L_{\rm 0.5-10\rm keV}$ from literatures
for the sources probably dominated by low-level accretion onto NSs. 
Specifically, we collect a sample composed of 16 non-pulsating NS-LMXBs with the measurements 
of $\eta$ and $L_{\rm 0.5-10\rm keV}$, 
i.e., Cen X-4 \citep[][]{Asai1996,Cackett2010,Cackett2013}, 
Aql X-1 \citep[][]{Rutledge2002a,Campana2014}, 
MXB 1659-29 \citep[][]{Wijnands2004c},
XTE J1709-267 \citep[][]{Jonker2004},
KS 1731-260 \citep[][]{Wijnands2001},
SAX J1810.8-2609 \citep[][]{Jonker2004b,Allen2018},
XTE J2123-058 \citep[][]{Tomsick2004}, 
4U 1608-52 \citep[][]{Asai1996}, 
EXO 1745-248  \citep[][]{Wijnands2005,RiveraSandoval2018},
AX J1754.2-2754, 1RXS J171824.2-402934 and 1RXH J173523.7-354013 \citep[][]{ArmasPadilla2013b}, 
XTE J1701-462 \citep[][]{Fridriksson2011},
Terzan 5 X-3 (Swift J174805.3-244637) \citep[][]{Bahramian2014}, 
EXO 0748-676 \citep[][]{Degenaar2009,Zhangguobao2011}, 
and GRS 1747-312 \citep[][]{Vats2018}. 
Meanwhile, we also include a typical accreting millisecond X-ray pulsar (AMXP)   
SAX J1808.4-3658 for comparison \citep[][]{Campana2002,Heinke2007,Heinke2009}. 
Some comments for the data are as follows.
KS 1731-260 and MXB 1659-29 are two quasi-persistent sources. The thermal soft X-ray component of 
the X-ray spectrum of KS 1731-260 and MXB 1659-29 has been explained as the emission of the
crust cooling \citep[][for review]{Wijnands2017}. However, the origin of the power-law component of 
the X-ray spectrum is unclear, which is very probably related with the low-level accretion onto NSs.
Here as a comparison, we still put the data of KS 1731-260 and MXB 1659-29 in the present paper. 
One can refer to Fig. \ref {f:fth_obs} for all the data in detail.  

As we can see in Fig. \ref {f:fth_obs}, for nearly all the data, the value of $\eta$ is greater than $\sim 10\%$.  
Meanwhile, it seems that systematically there is a positive correlation between $\eta$ and $L_{\rm 0.5-10\rm keV}$ for 
$L_{\rm 0.5-10\rm keV} \gtrsim 2\times 10^{33} \rm \ erg \ s^{-1}$, below which there is an anti-correlation 
between $\eta$ and $L_{\rm 0.5-10\rm keV}$ for some sources (see also in Fig. 5 of \citet[][]{Jonker2004}). 
The positive correlation between $\eta$ and $L_{\rm 0.5-10\rm keV}$ 
is dominated by the observational data of XTE J1709-267, in which $\eta$ decreases from $72\%$ to 
less than $19\%$ for $L_{\rm 0.5-10\rm keV}$ decreasing from $\sim 5\times 10^{34}$ to 
$2\times 10^{33} \rm \ erg \ s^{-1}$ \citep[][]{Jonker2004}.  
A possible anti-correlation between $\eta$ and $L_{\rm 0.5-10\rm keV}$ is observed in 
MXB 1659-29 in the range of $L_{\rm 0.5-10\rm keV} \sim 3\times 10^{32}-
3\times 10^{33} \rm \ erg \ s^{-1}$ \citep[][]{Wijnands2004c}, and in KS 1731-260 in the range of 
$L_{\rm 0.5-10\rm keV} \sim 2\times 10^{32}-2\times 10^{33} \rm \ erg \ s^{-1}$ \citep[][]{Wijnands2001}. 
However, we should keep in mind that, the derived value of $\eta$ for MXB 1659-29 and 
KS 1731-260 are only upper limits, the anti-correlation between $\eta$ and $L_{\rm 0.5-10\rm keV}$ is 
still to be confirmed by the observations in the future. 

In Fig. \ref {f:fth_obs}, we plot the theoretical results of $\eta$ as a function of $L_{\rm 0.5-10\rm keV}$ for 
different $f_{\rm th}$ with $\alpha=0.3$ as a comparison. The solid line is for $f_{\rm th}=1$, 
the dotted line is for $f_{\rm th}=0.1$, the dashed line is for $f_{\rm th}=0.05$, and the dot-dashed 
line is for $f_{\rm th}=0.01$. It is clear that, for nearly all the observational data, the value of  
$\eta$ is $\gtrsim 10\%$, which requires $f_{\rm th}\lesssim 0.1$.
As for EXO 1745-248, besides the data in \citep[][]{Wijnands2005}, the new observational data also 
show that the X-ray spectrum is nearly completely dominated by a power-law component in the range of 
$L_{\rm 0.5-10\rm keV} \sim 10^{32}-10^{34} \rm \ erg \ s^{-1}$, indicating an extreme value of 
$\eta \sim 100\%$ (only at the highest quiescent X-ray luminosity of 
$L_{\rm 0.5-10\rm keV} \sim 10^{34} \rm \ erg \ s^{-1}$, an additional thermal soft X-ray component is 
needed to improve the spectral fitting), the origin of which is still not well understood 
\citep[][]{RiveraSandoval2018}. If the power-law X-ray emission in EXO 1745-248 can be explained in the 
framework of our ADAF model, which means that $f_{\rm th}$ must be an extreme low value of 
$f_{\rm th} \sim 0$. 
SAX J1808.4-3658 is an AMXP, which could have a 
sizeable magnetic field (generally $\sim 10^{8-9} \rm G$). The quiescent spectrum of SAX J1808.4-3658 is 
completely dominated by a power-law component, the origin of which thus is probably related to the magnetic 
field \citep[e.g.][]{Stella1994,Burderi2003,Campana2002}. However, in the present paper SAX J1808.4-3658 is still 
included as in Fig. \ref {f:fth_obs}. This is because observationally the effect of the magnetic field on the 
X-ray spectrum of AMXPs is still under debate \citep[e.g.][]{Wijnands2015}. Actually, it is poorly understood 
why some NS-LMXBs show pulsations and others do not, and it is not proven that this is due to a difference in 
magnetic field strength. For instance, Aql X-1 is such a source with only the intermittent  
millisecond X-ray pulsations observed \citep[][]{Casella2008,Qiao2019}.
Again, if the power-law emission in SAX J1808.4-3658 can be explained with the ADAF model around a 
NS, which means that an extreme value of $f_{\rm th} \sim 0$ is required. 

In addition, as we can see in Fig. \ref {f:fth_obs}, our model intrinsically predicts a positive correlation
between $\eta$ and $L_{\rm 0.5-10\rm keV}$ for $L_{\rm 0.5-10\rm keV} \gtrsim$ a few times of  
$10^{33} \rm \ erg \ s^{-1}$, and an anti-correlation between $\eta$ and 
$L_{\rm 0.5-10\rm keV}$ for $L_{\rm 0.5-10\rm keV} \lesssim$ a few times of $10^{33} \rm \ erg \ s^{-1}$
for taking a relatively lower value of $f_{\rm th}=0.1, 0.05$ and $0.01$ respectively.
As can be seen that, the slope of both the positive correlation and the anti-correlation between 
$\eta$ and $L_{\rm 0.5-10\rm keV}$ can be significantly affected by the value of $f_{\rm th}$, i.e. 
both the positive correlation and the anti-correlation between $\eta$ and $L_{\rm 0.5-10\rm keV}$
become weaker with increasing $f_{\rm th}$. 
Here we expect that the predicted correlation between $\eta$ and $L_{\rm 0.5-10\rm keV}$ from our model
could be confirmed by more accurate observations in the future although there are some 
observational clues of such correlations, e.g., the observations for XTE J1709-267, MXB 1659-29 and KS 1731-260.  
We should also keep in mind that, as the first order approximation, currently, our theoretical results can only 
very roughly reproduce a trend of both the positive correlation and the anti-correlation. A more detailed 
calculation is still needed to more accurately explain the observations. Fox example, in this paper, 
we always assume that $f_{\rm th}$ is a constant for different $\dot m$. Actually, if we consider $f_{\rm th}$ 
as a function of $\dot m$, i.e. $f_{\rm th}(\dot m)$, definitely we can more flexibly fit the relation between 
$\eta$ and $L_{\rm 0.5-10\rm keV}$, which is beyond the scope of the present paper, 
and will be done in detail in the future work.

Finally, we would like to mention that in this paper, we set the NS mass $m=1.4$, and the NS radius 
$R_{*}=12.5$ km as in \citet[][]{Qiao2018b}. Theoretically, a change of $m$ and $R_{*}$ can affect the  
relation of $\eta$ versus $L_{\rm 0.5-10\rm keV}$ to some extent. However, observationally the value of  
$m$ and $R_{*}$ can be restricted in a very narrow range, i.e., $m\sim 1.4-2.0$ and $R_{*} \sim 10$ km.
\citep[e.g.][for review]{Degenaar2018b}. So a change of $m$ and $R_{*}$ in a reasonable range will not 
change our main conclusions. Meanwhile, as we know, when the LMXBs are in the hard state or the
quiescent state, it is generally suggested that the accretion flow exists in the form of a two-component structure 
with an inner ADAF plus an outer truncated thin disc, and the contribution of the truncated disc to the total 
emission depends on the truncation radius of the disc. The relative contribution of the truncated disc to the 
total emission increases with increasing the mass accretion rate as generally the truncation radius decreases 
with increasing the mass accretion rate. In particular, if the truncation radius is less than $\sim 10 R_{\rm S}$ 
(with $R_{\rm S}$ being the Schwarzschild radius, and $R_{\rm S}=2.95\times 10^{5}\ m\ \rm cm $), 
it is expected that the truncated disc can contribute a significant fraction of the thermal 
soft X-ray emission. However, theoretically only very fewer models can predict the value of the truncation radius 
for a given mass accretion rate. Here for example, the value of the truncation radius as a function of mass accretion 
rate has been proposed around a BH within the framework of the disc evaporation model 
\citep[e.g.][]{Liu1999,Meyer2000a,Meyer2000b,Qiao2009,Qiao2010,Qiaoetal2013,Taam2012}.
Specifically, in \citet[][]{Taam2012}, a formula for the truncation radius 
$r_{\rm tr}$ (in units of $R_{\rm S}$) as functions of $\dot m$, the viscosity parameter $\alpha$ and the 
magnetic parameter $\beta$ is given, which is as follows, 
\begin{eqnarray}\label{tr}
r_{\rm tr} \approx 17.3 \dot m^{-0.886} \alpha^{0.07} \beta^{4.61}.
\end{eqnarray}
It is clear that $r_{\rm tr}$ is very weakly dependent on $\alpha$, while  
is very strongly dependent on $\beta$. For example, if we take $\alpha=0.3$, $\beta=0.95$ for a weaker magnetic 
field as expected in ADAF, and $\dot m=5\times 10^{-3}$ (the upper limit that the ADAF solution can exist 
for $\alpha=0.3$), the truncation radius $r_{\rm tr}$ is $\sim 1370$, the contribution of which in 
soft X-ray emission can definitely be neglected. Here in the present paper as the first step we simply neglect 
the contribution of the truncated disc to the thermal soft X-ray emission and further to the value of $\eta$ if 
indeed the disc evaporation model works for the truncation of the accretion disc in NS-LMXBs.

\section{Discussions}
\subsection{The effect of the NS spin}\label{spin}
\begin{figure*}
\includegraphics[width=85mm,height=60mm,angle=0.0]{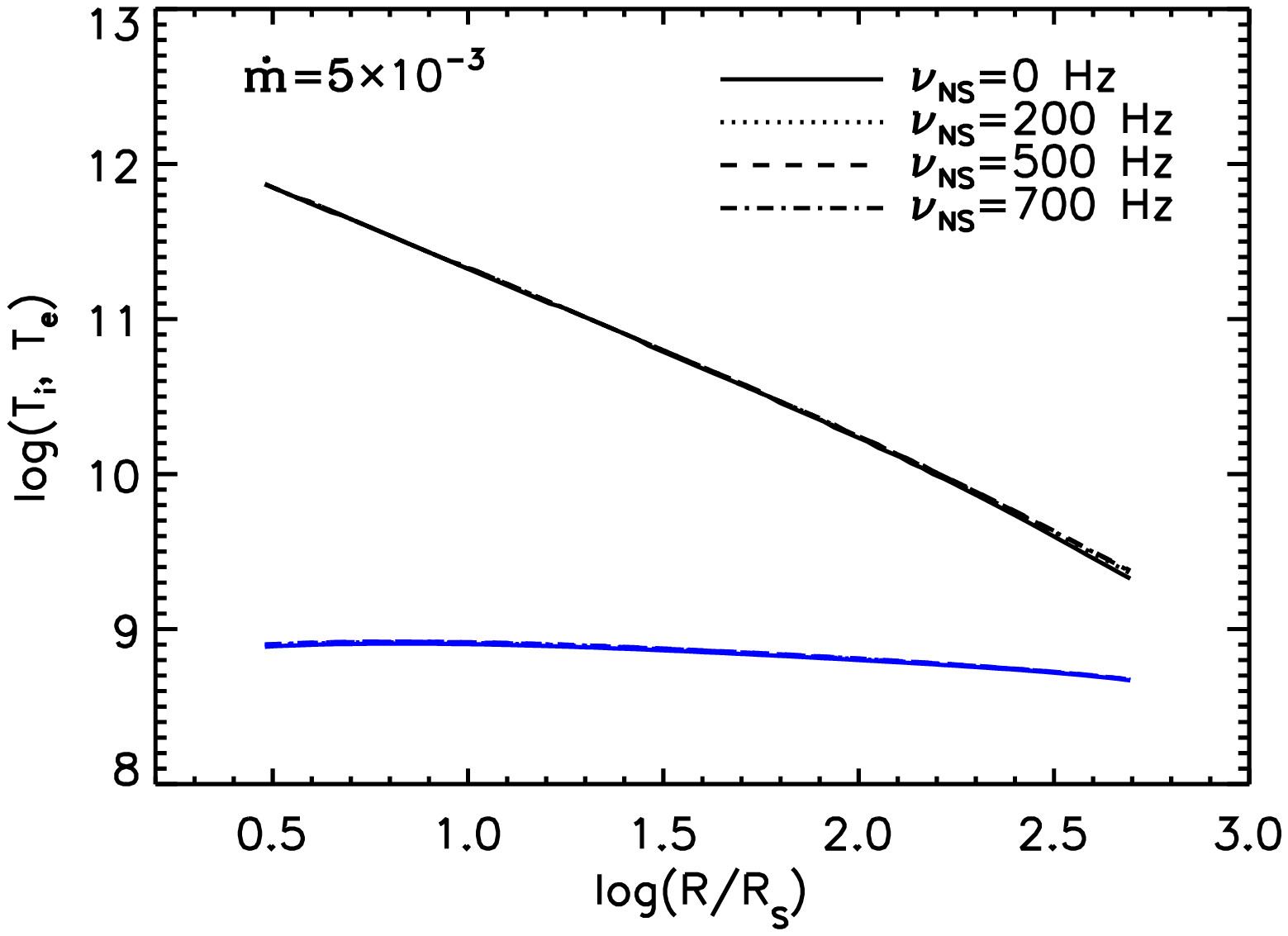}
\includegraphics[width=85mm,height=60mm,angle=0.0]{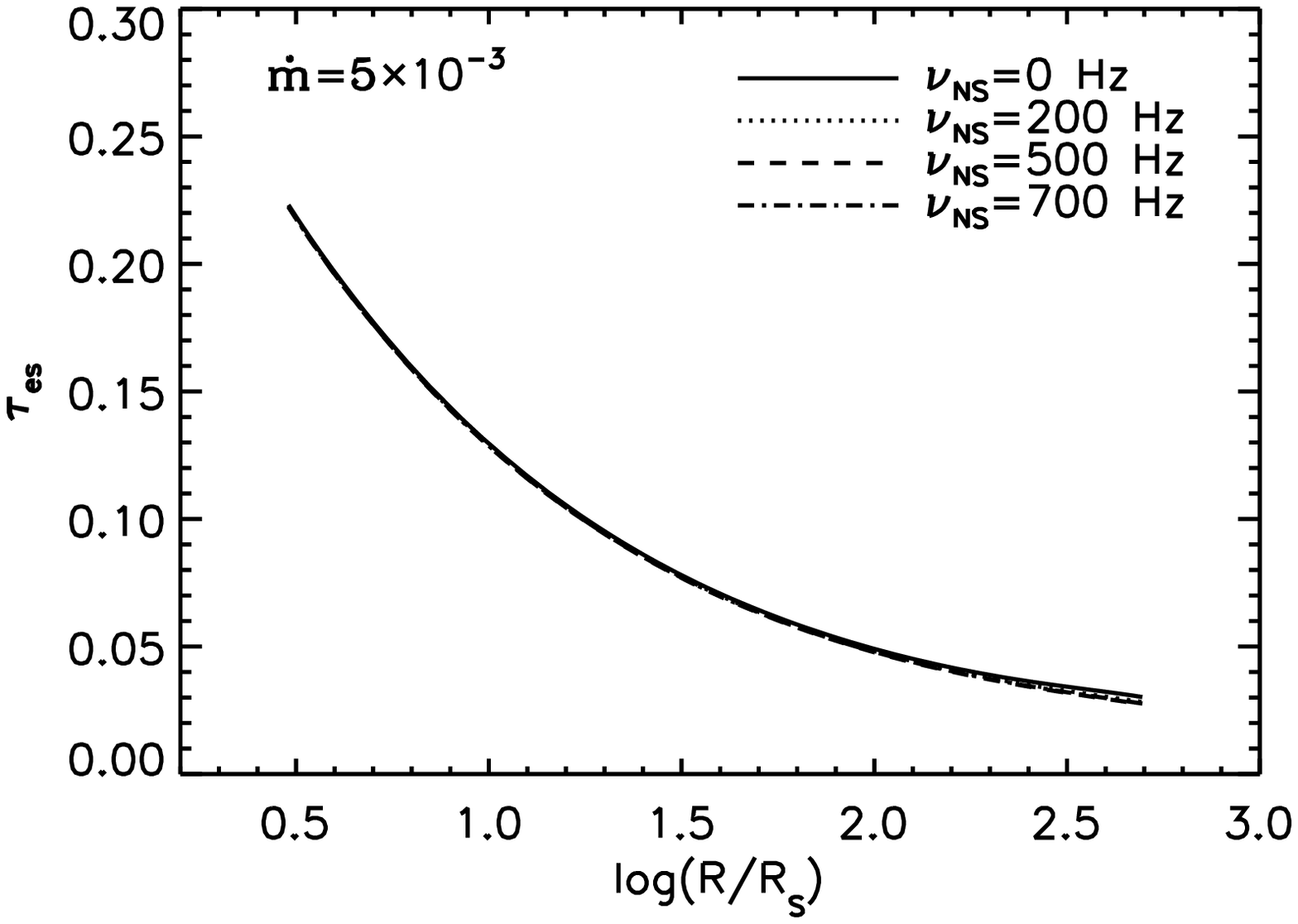}
\includegraphics[width=85mm,height=60mm,angle=0.0]{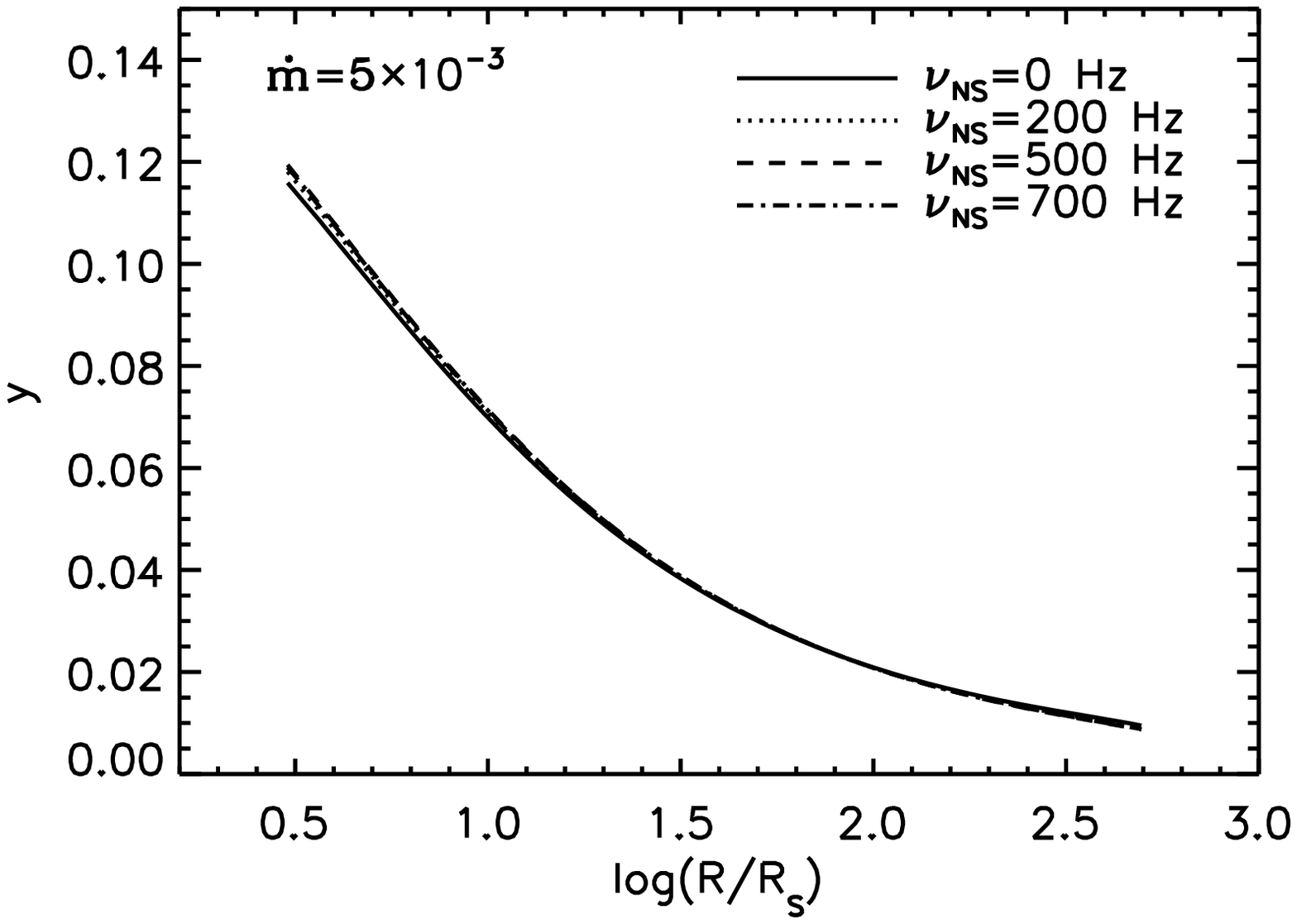}
\includegraphics[width=85mm,height=60mm,angle=0.0]{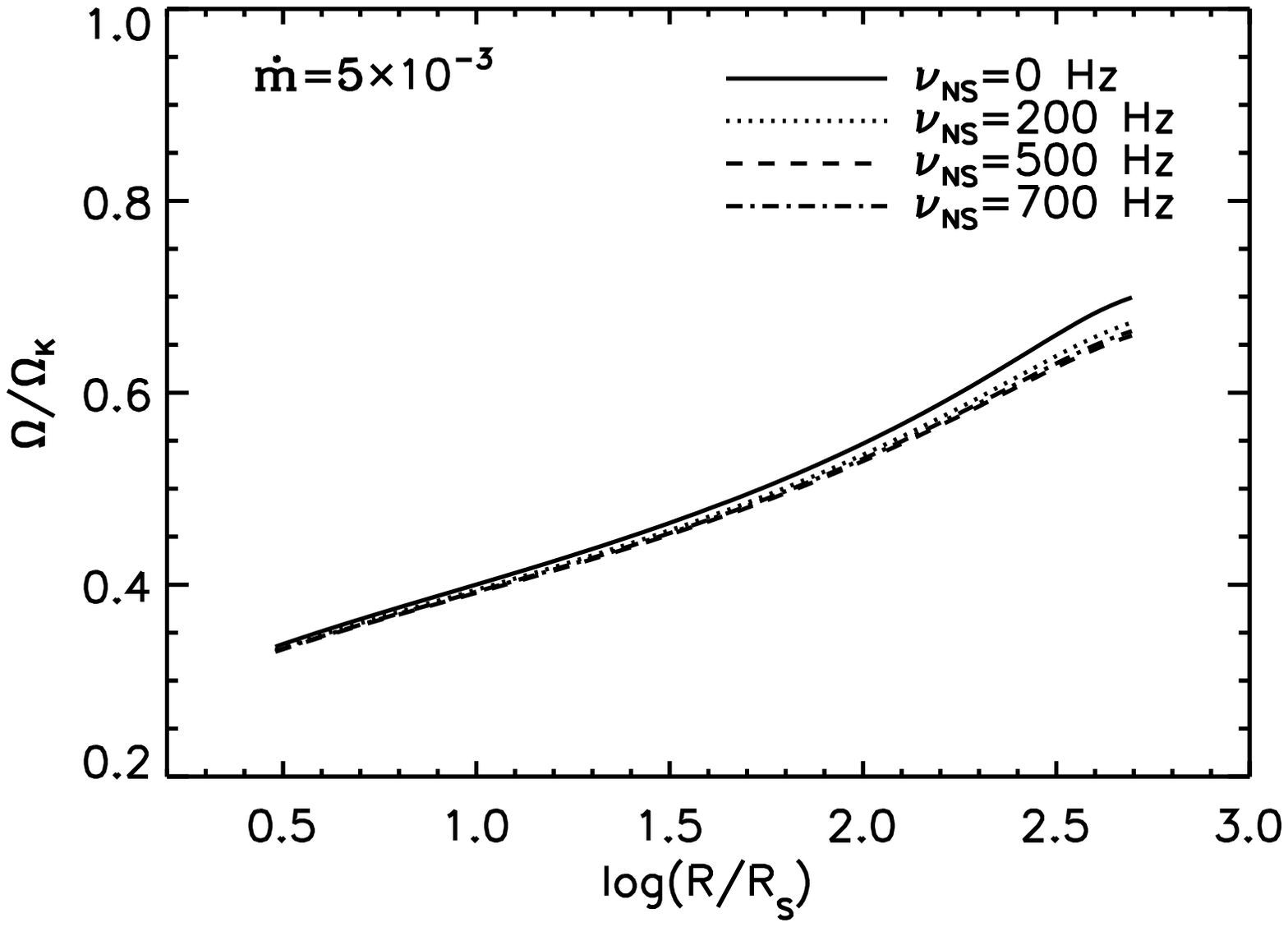}
\caption{\label{f:spin}
Panel (1): Ion temperature $T_{\rm i}$ (black line) and
electron temperature $T_{\rm e}$ (blue line) as a function of radius $R/R_{\rm S}$ around NSs for different 
rotational frequency of the NS $\nu_{\rm NS}$ with $\dot m=5\times 10^{-3}$.
Panel (2): Compton scattering optical depth $\tau_{\rm es}$ as a function of radius $R/R_{\rm S}$ around NSs
for different rotational frequency of the NS $\nu_{\rm NS}$ with $\dot m=5\times 10^{-3}$.
Panel (3): Compton $y$-parameter as a function of radius $R/R_{\rm S}$ around NSs for different 
rotational frequency of the NS $\nu_{\rm NS}$ with $\dot m=5\times 10^{-3}$.
Panel (4): $\Omega/\Omega_{\rm K}$ as a function of radius $R/R_{\rm S}$ around NSs 
for different rotational frequency of the NS $\nu_{\rm NS}$ with $\dot m=5\times 10^{-3}$.
In the calculation, we fix $\alpha=0.3$ and $f_{\rm th}=1.0$.}
\end{figure*}

In this paper, we calculate the structure and the corresponding emergent spectrum of the 
ADAF around a weakly magnetized NS within the framework of the self-similar solution of the ADAF 
as in \citet[][]{Qiao2018b}. Additionally, we consider the effect of the spin of the NS on the structure 
of the ADAF, which was not considered in \citet[][]{Qiao2018b}.
In the panel (1) of Fig. \ref{f:spin}, we plot the ion temperature $T_{\rm i}$ and the electron 
temperature $T_{\rm e}$ of the ADAF as a function of the radius $R/R_{\rm S}$ 
for different rotational frequency $\nu_{\rm NS}$ as taking $\nu_{\rm NS}=0, 200, 500$ and $700$ Hz 
respectively with $\dot m=5\times 10^{-3}$. In the calculation, we fix $\alpha=0.3$ and $f_{\rm th}=1.0$.
In the panel (2) of Fig. \ref{f:spin}, we plot the Compton scattering optical depth $\tau_{\rm es}$
in the vertical direction of the ADAF as a function of $R/R_{\rm S}$ 
for different $\nu_{\rm NS}$  as taking $\nu_{\rm NS}=0, 200, 500$ and $700$ Hz 
respectively with $\dot m=5\times 10^{-3}$.
In the panel (3) of Fig. \ref{f:spin}, we plot the Compton $y$-parameter of the ADAF as a function 
of $R/R_{\rm S}$ for different $\nu_{\rm NS}$  as taking $\nu_{\rm NS}=0, 200, 500$ and $700$ Hz 
respectively with $\dot m=5\times 10^{-3}$.
In the panel (4) of Fig. \ref{f:spin}, we plot the ratio of the angular velocity
of the ADAF to the Keplerian angular velocity, $\Omega/\Omega_{\rm K}$ 
(with $\Omega_{\rm K}$ being the Keplerian angular velocity), as a function 
of $R/R_{\rm S}$ for different $\nu_{\rm NS}$  as taking $\nu_{\rm NS}=0, 200, 500$ and $700$ Hz 
respectively with $\dot m=5\times 10^{-3}$.
It can be seen that the effects of $\nu_{\rm NS}$ on $T_{\rm i}$, $T_{\rm e}$, $\tau_{\rm es}$, $y$ and
$\Omega/\Omega_{\rm K}$ of the ADAF are very little, and can nearly be neglected. 
This is because, in this paper as in equation \ref {soft_2} of Section \ref {model}, 
the effect of $\nu_{\rm NS}$ is mainly reflected as the contribution of the soft photons  
to be scattered in the ADAF. However, the ADAF solution is hot, 
the energy of the ADAF transferred onto the surface of the NS is always dominated by the internal 
energy and the radial kinetic energy of the ADAF, i.e., $L_{*}^{'} \gg L_{\rm bl}$.
Since the effect of $\nu_{\rm NS}$ on the structure of the ADAF is very little, we expect 
that the effect of $\nu_{\rm NS}$ on the corresponding emergent spectrum is also very little, and
can be neglected. Here, we should keep in mind that in the present paper, we do not take 
into account the effect of general relativity for the different value of $\nu_{\rm NS}$ on the 
structure and the corresponding emergent spectrum of the ADAF, which is beyond the scope of 
the present paper, and will be studied in the future work. 

\subsection{On the physical origin of the X-ray emission in the quiescent state}\label{quiescent} 
In this paper, we theoretically investigate the relation of $\eta$ versus $L_{\rm 0.5-10\rm keV}$ in  
low-level accreting NSs ($L_{\rm 0.5-10\rm keV} \lesssim 10^{36}\ \rm erg \ s^{-1}$) within the framework of 
the self-similar solution of the ADAF around a weakly magnetized NS, in which the thermal soft X-ray component 
is from the surface of the NS, and the power-law component is from the ADAF itself.   
However, we would like to mention that when NS-LMXBs are in the quiescent state, i.e., 
$L_{\rm 0.5-10\rm keV}\lesssim 10^{34}\ \rm erg \ s^{-1}$, the origin of both the thermal soft X-ray 
component and the power-law component of the X-ray spectrum are still under debate \citep[e.g.][for review]{Wijnands2017}.  
Observationally, when NS-LMXBs go into the quiescent state, the X-ray spectra are diverse, and three 
possible X-ray spectra have been observed for different sources. The X-ray spectra can be, 
(1) totally dominated by the thermal soft X-ray component [hereafter spectrum (1)], 
or (2) totally dominated by the power-law component [hereafter spectrum (2)], 
or (3) described by the two-component model with a thermal soft X-ray component plus a power-law 
component [hereafter spectrum (3)].
The spectrum (1) could be produced by the emission of the crust cooling of the NS, which is heated   
during the outburst \citep[e.g.][]{Brown1998,Campana2000,Rutledge1999,Rutledge2001a,Rutledge2001b},
and also could be produced by the very low accretion on the surface of the NS, e.g, as the paper 
of \citet[][]{Qiao2018b}, or the present paper with an extreme value of $f_{\rm th}=1$ taken.
Currently, the origin of the spectrum (2) is still unclear. It has been proposed that such a spectrum 
might be related to the magnetic field of the NS as a pulsar-wind mechanism, such as the study of 
SAX J1808.4-3658 \citep[e.g.][]{Stella1994,Burderi2003,Campana2002}.
For the spectrum (3), the thermal soft X-ray component is possible to be produced by the crust cooling of 
the NS. Meanwhile, the study of the variability property of the thermal soft X-ray component indicates 
that the origin of the thermal soft X-ray component is more likely due to the low-level accretion onto the 
surface of the NS, such as the study of Cen X-4 \citep[][]{Campana1997,Campana2004,Cackett2010,Cackett2013}, 
and Aql X-1 \citep[][]{Rutledge2002a,Cackett2011,Bernardini2013}.
The origin of the power-law component in the spectrum (3) is unclear, and is also very possible to be 
related with the low-level accretion onto the surface of the NS \citep[][]{Zampieri1995,DAngelo2015,Chakrabarty2014}.
By fitting the quiescent spectrum of Cen X-4 ($L_{\rm 0.5-10\rm keV} \sim 10^{33}\ \rm erg \ s^{-1}$) 
with the joint $\it XMM-Newton$ and $\it Nuclear\ Spectroscopic\ Telescope\ Array\ (NuSTAR)$ observations, 
it is suggested that both the thermal soft X-ray component and the power-law component are due to the 
low-level accretion onto the surface of the NS \citep[][]{DAngelo2015}. 
Specifically, the authors fitted the thermal soft X-ray component with the \textsc{nsatmos} NS atmosphere 
model \citep{Heinke2006}, and found that the power-law component ($\Gamma\sim 1-1.5$) with a cut-off 
energy $\sim 10$ keV can be well fitted by the bremsstrahlung from the boundary layer between the surface 
of the NS and the accretion flow (with the accretion flow here being the radiatively inefficient accretion 
flow---RIAF). In \citet[][]{DAngelo2015}, the authors argued that the power-law emission comes from the boundary 
layer rather than the accretion flow itself, so the accretion flow as its name is `radiatively inefficient'. 
One can refer to Section \ref{efficiency} on this point for discussions. 
In a word, we think that the origin of the X-ray emission in the quiescent state of NS-LMXBs is very 
complicated, the understanding to which depends on both the strict theoretical and observational 
investigations in the future.  

\subsection{The radiative efficiency of NSs with an ADAF accretion}\label{efficiency}
The defining property of ADAF as its name `advection-dominated accretion flow', is that a large fraction of the 
viscously dissipated  energy will be stored in the accretion flow and advected onto the central compact 
object (BH or NS) rather than be directly radiated out in the accretion flow itself. In BH case, 
the advected energy will `vanish' as passing through the event horizon of the BH.  
While in NS case, it is often suggested that such energy advected onto the surface of the NS will 
eventually be radiated out \citep[][for review]{Narayan2008}.
As we can see in Section \ref {Sec:alpha}, the energy of the ADAF transferred onto the surface of the NS 
(including the internal energy, the radial kinetic energy and the rotational energy) is dependent on the 
viscosity parameter $\alpha$, i.e.,  $L_{*}/L_{\rm G}$ increases with increasing $\alpha$. The value of  
$L_{*}/L_{\rm G}$ is in the range of $\sim 60-80\%$ for different $\alpha$, which is higher than of the
thin disc of  $L_{*}/L_{\rm G}=50\%$ in the framework of the Newtonian mechanics. 
For the thin disc around a NS, half of the gravitational energy will be released in the accretion disc as, 
\begin{eqnarray}\label{e:disc}
L_{\rm d}={1\over 2}\dot M {GM\over R_{*}}, 
\end{eqnarray}
and the energy transferred onto the surface of the NS (or released in the boundary layer between the disc and the NS) 
can be seen as in equation \ref{soft_2}, and can be further re-expressed as, 
\begin{eqnarray}\label{e:BL}
L_{\rm bl}={1\over 2}{\dot M} k {GM\over R_{*}} \bigg(1-{\nu_{\rm NS}\over \nu_{*}}\bigg)^2, 
\end{eqnarray}
where $k$=1. The total radiative luminosity of the NS accreting system is $L_{\rm rad}=L_{\rm d}+L_{\rm bl}$.
If $\nu_{\rm NS}=0$, $L_{\rm d}$ is equal to $L_{\rm bl}$, and the radiative efficiency
of the NS accreting system is 
$\epsilon=L_{\rm rad}/{\dot Mc^2}= L_{\rm d}/{\dot Mc^2}+L_{\rm bl}/{\dot Mc^2} \sim 0.1+0.1=0.2$.

For the ADAF around a NS, if the energy transferred onto the surface of the NS can be 
eventually radiated out, i.e. taking $f_{\rm th}=1$ in our model, the total radiative luminosity 
is $L_{\rm rad}=L_{\rm ADAF}+L_{*}=\dot M {GM\over R_{*}}$, and the radiative efficiency of 
the NS accreting system is $\epsilon=L_{\rm rad}/{\dot Mc^2}\sim 0.2$. However, as can be seen in 
Section \ref{Sec:fth}, in order to match the observed range of the fractional contribution of 
the power-law luminosity $\eta$ and the relation of $\eta$ versus $L_{\rm 0.5-10\rm keV}$, 
it implies that $f_{\rm th}$ is at least less than $0.1$. For example, if we take 
$\alpha=0.3$ and $f_{\rm th}=0.1$, the total radiative luminosity is  
$L_{\rm rad}=L_{\rm ADAF}+f_{\rm th}L_{*} = L_{\rm ADAF}+0.1L_{*} \sim  
(100\%-60\%) \dot M {GM\over R_{*}} + 0.1*60\%*\dot M {GM\over R_{*}}= 0.46\dot M {GM\over R_{*}}$,
and the radiative efficiency is 
$\epsilon=L_{\rm rad}/{\dot Mc^2}\sim 0.46\times 2 \times 0.1\sim 0.09$.  
For $\alpha=0.3$ and $f_{\rm th}=0.01$, the result is similar, and the 
radiative efficiency is $\epsilon=L_{\rm rad}/{\dot Mc^2}\sim 0.406 \times 2 \times 0.1\sim 0.08$.  
Here we should note that, although the radiative efficiency of the NS may not be 
as high as the predicted results previously of $\epsilon \sim {\dot M GM\over R_{*}}/{\dot M c^2}\sim 0.2$,
the radiative efficiency of the NS with an ADAF accretion is still significantly higher that of the BH. 
The radiative efficiency of the NS with an ADAF accretion nearly does not depend on $\dot m$ with 
$\epsilon$ always $\sim 0.08-0.09$. The radiative efficiency of the BH with an ADAF accretion
is completely different, which is very dependent on $\dot m$. If $\dot m$ is approaching the critical mass accretion rate of 
$\dot M_{\rm crit}^{'} \sim \alpha^2 \dot M_{\rm Edd}$, the radiative efficiency is approaching the 
value of the thin disc as $\epsilon\sim 0.1$, while 
$\epsilon$ dramatically decreases with decreasing $\dot m$ with a form of $\epsilon \propto m^{a}$.
Specifically, $a$ is $0.71$ for $\dot m\lesssim 7.6\times 10^{-5}$, $a$ is 0.47  for 
$7.6\times 10^{-5}\lesssim  \dot m \lesssim 4.5\times 10^{-3}$, and 
$a$ is $3.67$ for $4.5\times 10^{-3}\lesssim  \dot m \lesssim 7.1\times 10^{-3}$ (with the parameter 
of direct heating to the electrons $\delta=10^{-3}$ assumed). For example, 
$\epsilon \sim 8\%$, $\sim 0.6\%$, $\sim 0.2\%$ and $\sim 0.04\%$ for 
$\dot m =10^{-2}$, $10^{-3}$, $10^{-4}$ and $10^{-5}$ respectively \citep[][]{Xie2012}.

In this paper,  we assume that only a fraction, $f_{\rm th}$, of the total energy of the ADAF  
transferred onto the surface of the NS is thermalized at the surface of the NS as the soft photons to be
scattered in the ADAF itself, which means that the remaining fraction, 1-$f_{\rm th}$, of the total energy 
of the ADAF transferred onto the surface of the NS
is not radiated out. One of the possibilities is that such a fraction of the energy is converted to the 
rotational energy of the NS, making the rotational frequency $\nu_{\rm NS}$ of the NS increased.  
For example, if we take $\alpha=0.3$ and $f_{\rm th}=0.1$, the energy converted to  
the rotational energy of the NS per second is 
$\dot E=(1-f_{\rm th})L_{*}=(1-0.1)*60\%*\dot M {GM\over R_{*}}\sim 0.5\dot M {GM\over R_{*}}$. 
For a typical value of $M=1.4 M_{\odot}$, $R_{*}=12.5\ \rm km$, and $\dot M=10^{-3}\dot M_{\rm Edd}$, 
$\dot E \sim 1.76\times 10^{35} \rm \ erg \ s^{-1}$. As we know, the rotational kinetic energy of 
the NS can be expressed as, 
\begin{eqnarray}\label{e:rot}
E={1\over 2}I\Omega_{\rm NS}^2, 
\end{eqnarray}
where $I$ is the moment of inertia, and $\Omega_{\rm NS}$ (with $\Omega_{\rm NS}=2\pi\nu_{\rm NS}$) is the 
rotational angular velocity of the NS.  For a typical value of $M=1.4 M_{\odot}$, and 
$R_{*}=12.5\ \rm km$, $I \sim 1.75\times 10^{45}\rm g\ cm^2$. Assuming $\nu_{\rm NS}=500$ Hz, 
$E\sim 8.7\times 10^{51} \ \rm erg$. Roughly, the time scale for the rotational frequency of the NS increases 
from $\nu_{\rm NS}=0$ to $\nu_{\rm NS}=500$ is $t\sim E/\dot E\sim 1.6\times 10^9$ years. So a significant 
change of the rotational frequency in the range of our observable time is very little, and nearly can be neglected.
Meanwhile, more importantly, as discussed in Section \ref {spin}, the effect of the NS spin
on the structure and corresponding emergent spectrum is very little, so our assumption in this paper 
is self-consistent.  We should also keep in mind, in the present paper, all our conclusions are 
based on the self-similar solution of the ADAF around a weakly magnetized NS. Especially, we are not clear   
whether the structure and the radiative property  of the ADAF can significantly change or not for 
different $\nu_{\rm NS}$ if the global solution of the ADAF around a NS is considered, which is very 
important and will be studied in the future work.   

Here we would like to mention an interesting paper of \citet[][]{DAngelo2015}. 
As has been discussed in Section \ref {quiescent}, by fitting the broad band quiescent X-ray spectrum 
of Cen X-4, \citet[][]{DAngelo2015} concluded that the power-law component of the X-ray spectrum is from 
the boundary layer between the surface of the NS and the accretion flow rather than the accretion flow itself, 
and further concluded that the accretion flow is `radiatively inefficient' as similar as 
\citet[][]{Chakrabarty2014}. However, as far as the viewing point of the energy balance, in \citet[][]{DAngelo2015}, 
although the accretion flow is `radiatively inefficient', most of the energy released in the accretion 
process is radiated out in the boundary layer, the whole accreting system itself is still radiatively efficient.
Actually, the radiative efficiency from \citet[][]{DAngelo2015} is not deviated from that of our model. 
Furthermore, we would like to point out that the fraction of the accretion energy released in the boundary 
layer or the accretion flow itself is uncertain, which can be affected by several factors, such as the spin and 
the magnetic field of the NS. A more detailed study of the low-level accretion onto NSs by considering the 
spin and the magnetic field of the NS is still necessary in the future work.

\section{Conclusions}
In this paper, we systematically investigate the X-ray emission in weakly magnetized low-level accreting NSs within the 
framework of the self-similar solution of the ADAF as in \citet[][]{Qiao2018b}, in which the soft X-ray 
emission is from the surface of the NS and the power-law X-ray emission is from the ADAF itself respectively.
Additionally, in this paper, we update the calculation of \citet[][]{Qiao2018b} with the effect of the NS spin 
considered. We test the effect of the viscosity parameter $\alpha$ and the thermalized  parameter $f_{\rm th}$ 
on the relation between the fractional contribution of the power-law luminosity 
$\eta$ and the X-ray luminosity $L_{\rm 0.5-10\rm kev}$.
It is found that the effect of $\alpha$ on the relation  between $\eta$ and $L_{\rm 0.5-10\rm kev}$ is very little,
and nearly can be neglected, while the effect of $f_{\rm th}$ on the relation between $\eta$ and $L_{\rm 0.5-10\rm kev}$ 
is very significant. By comparing with a sample of non-pulsating NS-LMXBs probably dominated by low-level 
accretion onto NSs, it is found that a small value of $f_{\rm th}\lesssim 0.1$ is needed to match the observed 
range of $\eta \gtrsim 10\%$ in the diagram of $\eta$ versus $L_{\rm 0.5-10\rm keV}$.
We argue that the small value of $f_{\rm th}\lesssim 0.1$ implies that 
only a small fraction of the energy transferred onto the surface of the NS can be radiated out for the ADAF 
around a NS. Meanwhile, we think that it is very possible that the remaining fraction, 1-$f_{\rm th}$, of the 
energy transferred onto the surface of the NS can be converted to the rotational kinetic energy of the NS. 
So the radiative efficiency of a weakly magnetized NS with an ADAF accretion may not be as higher as the predicted 
results previously of $\epsilon \sim 0.2$ despite the existence of the hard surface.   
Here we would like to mention that although a great of efforts have been made for   
investigating the X-ray spectrum of NS-LMXBs as in the low-level accretion regime, the detailed physical 
mechanism for the interaction between the accretion flow and the NS is still under debate 
\citep[e.g.][for discussions]{Wijnands2015}. 
Finally, our model intrinsically predicts a positive correlation 
between $\eta$ and  $L_{\rm 0.5-10\rm keV}$ for $L_{\rm 0.5-10\rm keV} \gtrsim$ a few times of  
$10^{33} \rm \ erg \ s^{-1}$, and an anti-correlation between $\eta$ and 
$L_{\rm 0.5-10\rm keV}$ for $L_{\rm 0.5-10\rm keV} \lesssim$ a few times of $10^{33} \rm \ erg \ s^{-1}$
for taking $f_{\rm th}=0.1, 0.05$ and $0.01$ respectively, which we expect can be confirmed by 
more accurate observations in the future. 
\section*{Acknowledgments}
This work is supported by the National Natural Science Foundation of
China (Grants 11773037 and 11673026), the gravitational wave pilot B (Grants No. XDB23040100), 
the Strategic Pioneer Program on Space Science, Chinese Academy of Sciences (Grant  No.
XDA15052100) and the National Program on Key Research and Development Project 
(Grant No. 2016YFA0400804).

\bibliographystyle{mnras}
\bibliography{qiaoel}


\bsp	
\label{lastpage}
\end{document}